\documentclass{ieeeaccess}
\usepackage{cite}
\usepackage{amsmath,amssymb,amsfonts}
\usepackage{algorithmic}
\usepackage{graphicx}
\usepackage{textcomp}
\usepackage{color, colortbl}
\usepackage{adjustbox}
\usepackage{booktabs}
\usepackage{tabularx}
\usepackage{subcaption}
\usepackage{makecell}

\usepackage{array}
\usepackage{caption}
\usepackage{multirow}
\usepackage{url}
\usepackage{wrapfig}
\usepackage{float}
\usepackage{hyperref}
\usepackage{adjustbox}
\usepackage{rotating}
\usepackage{array}
\usepackage{makecell}
\usepackage{afterpage}
\usepackage{supertabular}
\usepackage{longtable,lscape}
\usepackage{multirow}
\usepackage{enumitem}
\usepackage[para,online,flushleft]{threeparttable}
\hypersetup{colorlinks, citecolor={black}, filecolor=black, linkcolor=black, urlcolor=black}
\definecolor{Gray}{gray}{0.9}
\definecolor{LightGray}{gray}{0.6}
\definecolor{red}{rgb}{0, 0, 0}

\usepackage{xcolor}
\definecolor{green(munsell)}{rgb}{0.0, 0.66, 0.47}
\definecolor{cadmiumgreen}{rgb}{0.0, 0.42, 0.24}
\definecolor{cobalt}{rgb}{0.0, 0.28, 0.67}
\definecolor{amber(sae/ece)}{rgb}{1.0, 0.49, 0.0}
\newlength\MAX  
\newcommand*\Chart[3]{
\rlap{
    \textcolor{black!30}{
        \rule{\MAX}{2ex}
        }
    }
\textcolor{cobalt}{\rule{#1\MAX}{2ex}}\textcolor{green(munsell)}{\rule{#2\MAX}{2ex}}\textcolor{amber(sae/ece)}{\rule{#3\MAX}{2ex}}
}

\newlength\BARSIZE  
\newcommand*\ChartBarBlue[1]{\textcolor{cobalt}{\rule{\BARSIZE}{2ex}}}
\newcommand*\ChartBarGreen[1]{\textcolor{green(munsell)}{\rule{\BARSIZE}{2ex}}}
\newcommand*\ChartBarOrange[1]{\textcolor{amber(sae/ece)}{\rule{\BARSIZE}{2ex}}}

\def\BibTeX{{\rm B\kern-.05em{\sc i\kern-.025em b}\kern-.08em
    T\kern-.1667em\lower.7ex\hbox{E}\kern-.125emX}}
\begin{document}
\history{Date of publication May 30, 2022, date of current version May 30, 2022.}
\doi{10.1109/ACCESS.2022.DOI}

\title{Sensing Eating Events in Context: A Smartphone-Only Approach}

\author{
\uppercase{Wageesha Bangamuarachchi}\authorrefmark{1,*},
\uppercase{Anju Chamantha}\authorrefmark{1,*},
\uppercase{Lakmal Meegahapola}\authorrefmark{2,3,*},
\uppercase{Salvador Ruiz-Correa}\authorrefmark{4},
\uppercase{Indika Perera}\authorrefmark{1},
\uppercase{Daniel Gatica-Perez\authorrefmark{2,3}}}

\address[1]{University of Moratuwa, Moratuwa 10400, Sri Lanka}
\address[2]{Idiap Research Institute, Martigny 1920, Switzerland}
\address[3]{\'Ecole Polytechnique F\'ed\'erale de Lausanne (EPFL), Lausanne 1015, Switzerland}
\address[4]{IPICYT, San Luis Potosí S.L.P. 78216, Mexico}
\address[*]{Co-Primary, Listed Alphabetically}
\tfootnote{This work was funded by the European Commission’s Horizon 2020 project "WeNet: The Internet of Us", under grant agreement 823783.}

\markboth
{Bangamuarachchi, Chamantha, and Meegahapola et al.}
{Sensing Eating Events in Context: A Smartphone-Only Approach}

\corresp{Corresponding author: Lakmal Meegahapola (e-mail: lakmal.meegahapola@epfl.ch)}

\begin{abstract}

While the task of automatically detecting eating events has been examined in prior work using various wearable devices, the use of smartphones as standalone devices to infer eating events remains an open issue. This paper proposes a framework that infers eating vs. non-eating events from passive smartphone sensing and evaluates it on a dataset of 58 college students. First, we show that time of the day and features from modalities such as screen usage, accelerometer, app usage, and location are indicative of eating and non-eating events. Then, we show that eating events can be inferred with an AUROC (area under the receiver operating characteristics curve) of 0.65 using subject-independent machine learning models, which can be further improved up to 0.81 for subject-dependent and 0.81 for hybrid models using personalization techniques. Moreover, we show that users have different behavioral and contextual routines around eating episodes requiring specific feature groups to train fully personalized models. These findings are of potential value for future mobile food diary apps that are context-aware by enabling scalable sensing-based eating studies using only smartphones;  detecting under-reported eating events, thus increasing data quality in self report-based studies; providing functionality to track food consumption and generate reminders for on-time collection of food diaries; and supporting mobile interventions towards healthy eating practices. 
\end{abstract}


\titlepgskip=-15pt

\maketitle

\section{Introduction}\label{sec:introduction}

According to prior work in nutrition science and public health, unhealthy eating practices could lead to severe conditions such as heart disease, diabetes, high blood pressure, and high cholesterol \cite{Santani2018, Rahman2016, bedri2017earbit}. Hence, understanding the etiology and managing eating behavior is crucial. Fueled by such motivations, researchers have come up with different techniques to detect and monitor food intake, among which keeping food diaries (also known as food journaling and food logging) is one of the most common ones \cite{jung2020foundations}. On an individual level, food diaries help users with self-awareness, self-monitoring, and behavior change, and have also helped people with weight loss goals \cite{ ye2016assisting, jung2020foundations}. At the population level, they help researchers to conduct large-scale studies to understand population-level food consumption \cite{jung2020foundations}. Food diaries originated as a pencil-and-paper based technique \cite{alexandercan}, but in recent times mobile food diaries have become popular, and widely adopted commercial mobile health (mHealth) apps such as Samsung Health \cite{SamsungHealth2021} and MyFitnessPal \cite{MyFitnessPal2021} allow users to keep food diaries and facilitate mindful eating. 

While keeping a food diary has many benefits, it is difficult to sustain the practice of reporting all food intake over long periods due to a plethora of personal, societal, and technological factors such as forgetting to report food, losing motivation to report, and self and recall biases (e.g,, not reporting all eating events intentionally) \cite{Meegahapola2021OneMoreBite, jung2020foundations, ye2016assisting}. Such drawbacks have called for tools and techniques to automatically recognize eating events, as this would allow reminding users to report food intake on time. Prior studies in mobile sensing have used wrist wearables \cite{thomaz2015practical, ye2016assisting, thomaz2017exploring}, jaw-bone wearables \cite{san2020eating}, earables \cite{bi2018auracle, bedri2017earbit}, necklaces \cite{chun2018detecting, zhang2020necksense}, and other sensing modalities \cite{rahman2016predicting, merck2016multimodality} to detect eating by sensing wrist movements, bites, swallowing, and mastication among many other actions. While most of these techniques have shown promising performance in lab settings, some have also performed reasonably in everyday life conditions. However, these techniques require specific hardware configurations and wearables to be worn, which might be both a hassle for some users and unaffordable for others. Furthermore, wearable-based eating detection systems would have to maintain a connection with smartphones to automatically trigger actions on the phone, which requires bluetooth, wifi, or data connections to be kept turned on. As wearables are known for low battery life, the need to run continuously could drain the battery even faster. In contrast, recognizing eating events directly on the smartphone could address some of these usability and technical issues. Moreover, unlike wearables, smartphone coverage and mHealth app usage are already high in many countries \cite{Mobius2021}. For example, 96\% of young adults aged 18-29 in the United States own a smartphone \cite{PewResearchCentre2021}, and Nutrition and Diet apps has become the second most common app category among mHealth app users, just behind Fitness apps \cite{Mobius2021}. Prior studies in mobile sensing have looked into improving mobile food diaries with context-awareness \cite{Meegahapola2021OneMoreBite, Biel2018, Meegahapola2020Alone, Meegahapola2020Protecting}. However, whether smartphones alone be used to recognize eating events remains an open question. Considering these aspects, sensing eating events on smartphones could provide the following benefits:

\textit{Automatic Food Intake Tracking and Reminders:} Keeping mobile food diaries manually can be cumbersome as people tend to forget to report \cite{bell2020automatic}. Automatic eating event recognition could, on the one hand, keep track of eating events to provide feedback and remind users to report forgotten eating events. If such inferences were combined with other inferences such as food type \cite{Biel2018}, social context \cite{Meegahapola2020Alone}, or food consumption level \cite{Meegahapola2021OneMoreBite}, a holistic food diary could be maintained with minimal user input. This vision has been discussed in prior dietary monitoring \cite{bell2020automatic} and sensing \cite{Meegahapola2021Survey} literature.
    
\textit{Self-Report Validation:} A challenge in self-report-based questionnaires is under-reporting, i.e., participants failing to report eating events \cite{bell2020automatic}. In the context of mobile food diaries, when an actual eating event is incorrectly considered as a non-eating event, it adds noise to the data, which can be detrimental when training machine learning models. A low-cost smartphone-based system that could estimate if a period contained an actual eating or non-eating moment could filter out highly confident self-reports from noisy reports. This would lead to higher quality labels for public health studies and for training machine learning models in research and commercially available food diaries. 
    
\textit{Mobile Interventions: }After determining whether someone is eating or not, many subsequent inferences to determine the behavior and context of eating could be made \cite{Biel2018, Meegahapola2020Protecting, Meegahapola2021OneMoreBite, Meegahapola2020Alone}. These inferences  could help provide context-aware interventions and feedback to app users \cite{bell2020automatic}. More importantly, sensing eating events with the phone could be done without relying on external wearables, possibly reaching larger and more diverse populations.
    
\textit{Population-Level Studies:} Smartphone-only inferences could make it easier for nutrition scientists and dietitians to conduct population-level eating studies among larger populations, without the need for additional hardware such as wearable devices. Current population-level eating-related studies are predominantly done using self-reports. Some drawbacks of such studies, related to participant burden and attention limits of self-reports, could be addressed by automatically detecting eating events. Prior work has discussed the use of wearables for automatic dietary monitoring in population-level studies \cite{bell2020automatic}. Furthermore, a recent study discussed the detection of eating events using wearables to trigger further data capture on mobile food diaries \cite{Morshed2020}. However, conducting large-scale studies using additional wearable devices is expensive. Hence, the proposed method, based solely on smartphone sensing, could be helpful for nutrition researchers and dietitians to overcome issues in current methods, and facilitate the implementation of population-level studies by automatically inferring eating events.

In summary, while the characterization of eating has been attempted in the smartphone sensing domain (social context \cite{Meegahapola2020Alone}, food consumption level \cite{Meegahapola2021OneMoreBite}, food category and type \cite{Biel2018}), the use of smartphones as standalone devices to infer eating events remains as an open issue. In this paper, we examine whether smartphone sensing features could be used to classify time windows as corresponding to eating vs. non-eating events using subject-dependent, hybrid, and subject-independent machine learning models \cite{ferrari2020personalization}. 

We pose two research questions:

\noindent \textbf{RQ1:} What situational contexts and behaviors around eating and non-eating events can be observed by analyzing everyday eating events of a group of college students obtained via passive smartphone sensing?
\\
\noindent \textbf{RQ2:} Can eating and non-eating events be inferred by only using passive smartphone sensing?

By addressing the above research questions, this paper provides the following contributions: 

\noindent \textbf{Contribution 1:} We analyzed how passive smartphone sensing features differ for eating and non-eating in everyday life situations. As a case study, a dataset of over 12000 events provided by 58 college students in Mexico is used. We show that features from modalities such as application usage, accelerometer, location, and time of the day are informative of eating and non-eating events.

\noindent \textbf{Contribution 2:} We define and evaluate the task of inferring eating and non-eating events using passive smartphone sensing data, obtaining an AUROC of 0.65 with subject-independent models, which can be increased to 0.81 with subject-dependent models. Moreover, we show that feature selection plays a key role when training subject-dependent models. Each user might need models that use different features (compared to others) to achieve high performance. This shows the behavioral diversity of people around eating and the need to consider such diversity in building machine learning models that recognize eating events. We also found that hybrid models (partially personalized) perform reasonably well and on par with subject-dependent models. The results illustrate the potential of using passive smartphone sensing for building context-aware and automated mobile food diaries.

This paper is organized as follows. Section~\ref{sub:related_work}, describes the background and discusses related work. In Section~\ref{sec:mobile_app}, the study design, data collection procedure, and feature extraction techniques are provided. Section~\ref{sec:data_analysis}, presents a descriptive analysis and a statistical analysis of the dataset. The inference task is defined and evaluated in Section~\ref{sec:inference}. A number of important issues are discussed in Section~\ref{sec:discussion}. Finally, the paper is concluded in Section~\ref{sec:conclusion}. 

%\newpage 

\section{Background and Related Work}\label{sub:related_work}

\subsection{Nutrition Science Perspective}\label{subsec:context_as_a_proxy}

\paragraph{Eating as a Holistic Event}
Guided by how and why people eat, Bisogni et al. \cite{bisogni2007dimensions} provided a contextual framework for eating and drinking events, describing them as holistic events with eight interconnected dimensions. The dimensions are food and drink (type, amount, source, how consumed), time (chronological, relative experienced), location (general/specific, food access, weather/temperature), activities (nature, salience, active or sedentary), social setting (people present, social processes), mental processes (goals, emotions), physical condition (nourishment, other status), and recurrence (commonness, frequency, what recurs). The primary idea behind this framework is that situational and behavioral factors guide eating. Further, Jastran et al. \cite{jastran2009eating} showed that eating routines are embedded in the schedules around daily lives related to family, work, and recreation. They also said that repetitive patterns could be found in eating episodes among participants regarding the type of food and the situational context in which food was consumed. Similar ideas were proposed in other studies that also showed that social context, activity levels and types, psychological aspects, location, food availability, and several other situational and behavioral factors could affect the eating behavior \cite{Minati2014, Herman2005}. Ma et al. \cite{ma2015food} found that people in small towns and rural areas tend to travel sizable distances for lunch. In larger towns, people consume lunch at places closer to their workplace (e.g., canteen, close by restaurants, fast-food outlets) because of the limited time available for eating. This shows how eating behavior is related to dimensions such as recurrence, location, and time. Further, some studies examined links between app usage and eating behavior. For example, Turner et al. \cite{turner2017instagram} showed that excessive Instagram usage could be indicative of Orthorexia Nervosa, a condition of having an obsession with maintaining a healthy eating behavior, including a focus on healthy eating, food anxiety, and dietary restrictions. 

Even though not conclusive, such studies point towards modern mobile social media playing a role in shaping eating behavior. Hence, all these nutrition and behavioral sciences studies demonstrate how different behaviors and situational factors could affect eating behavior. In addition, they also show the repetitive nature of different dimensions around eating events, which could be helpful in terms of modeling eating behavior using smartphone sensing and machine learning. Moreover, even though there is no concrete definition regarding eating events or episodes \cite{bell2020automatic}, going with the terminology regarding holistic eating behavior, throughout this paper, the term \textit{\textbf{eating episode}} is used as the actual time period of food intake. Moreover, the time period around the food intake that also contains behaviors and contexts around the eating episode (e.g., going to the place of eating and coming back, using particular mobile apps before/after/while eating, etc.), that help us to consider eating as a holistic event is termed as \textit{\textbf{eating event}}. Hence, an eating event is the eating episode, and behavior and context before and after the eating episode captured with a time window. This definition is in line with prior work in mobile sensing that looked into characterizing eating and drinking events \cite{Biel2018, Bae2017, Meegahapola2021OneMoreBite, Meegahapola2020Alone}. More details regarding how these terms are operationalized can be found in Section~\ref{subsec:eatingepisode}.

% citation - kim2011eating
% Title - Eating patterns and nutritional characteristics associated with sleep duration
% conclusion - Disrupted eating patterns and diet of poor nutritional quality may exacerbate the development of obesity and metabolic diseases in habitual short and very long sleepers.

% citation - bellissimo2007effect
% title - Effect of television viewing at mealtime on food intake after a glucose preload in boys
% conclusion - TV viewing by boys while eating a meal contributes to increased energy intake and it does so by delaying normal mealtime satiation and reducing satiety signals from previously consumed foods

% citation - marshall2003meal
% title - Meal construction: exploring the relationship between eating occasion and location
% conclusion - meal occasion and eating location are not distinct elements but are intricately linked when examining food choice combinations, meal construction and acceptability. Differences in food choices between two groups (australian and british) may indeed be mediated by the perceptions of meal occasions and eating locations

\begin{figure*}[t]
\begin{center}
    \begin{minipage}[t]{\textwidth}
        \centering
        \includegraphics[width=\textwidth]{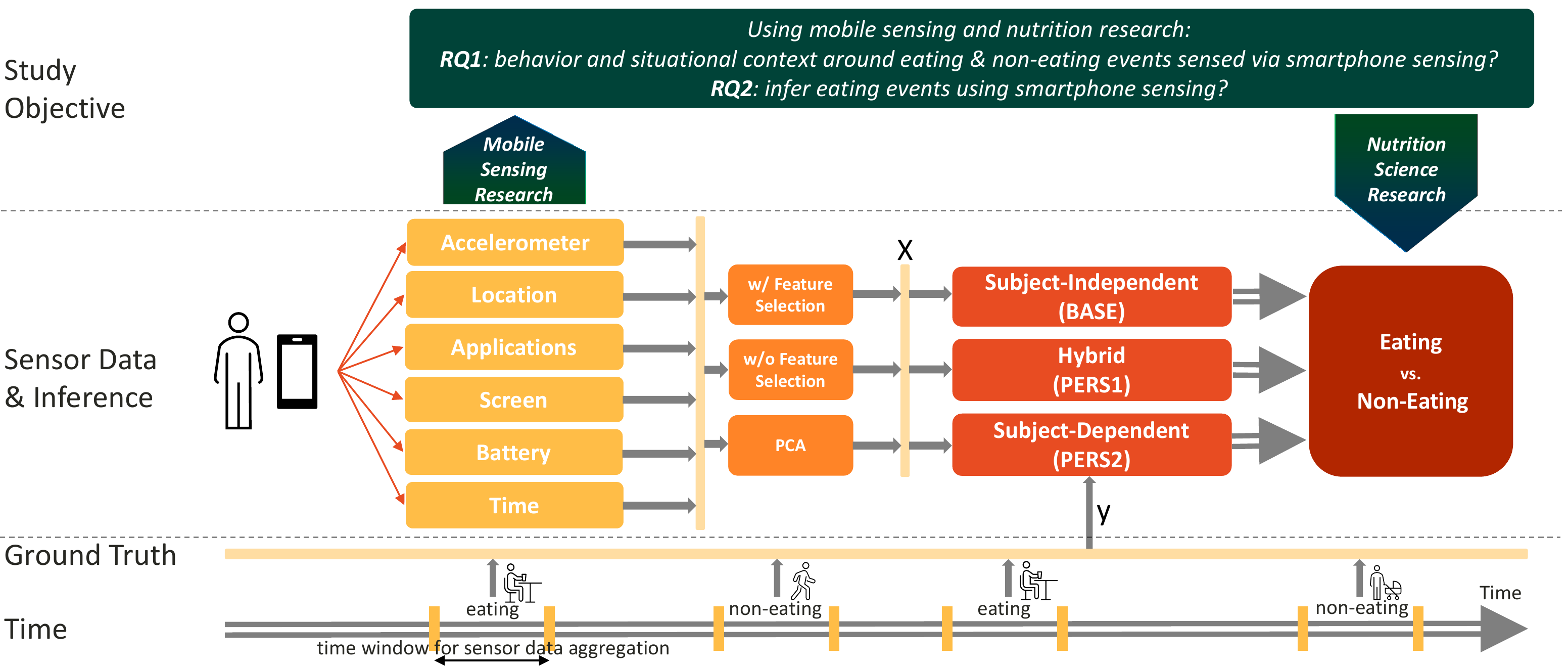}
        \caption{Objective of the Study}
        \label{fig:objective}
    \end{minipage}
\end{center}
% \vspace{-0.3 in}
\end{figure*}

\paragraph{Situational Context and Behavior as Proxies to Eating Events}Mobile sensing studies collect passive sensing and self-report data that can be broadly categorized into three pillars \cite{Meegahapola2021Survey}: person, behavior, and context. What this means is that each sensing modality will be taken as a proxy to a trait that is related to a person (i.e., mood, stress, sociability, age, sex, etc.), behavior (i.e., activities, routines, etc.), or context (i.e., location, social context, environmental context, date and time, etc.). Prior work in smartphone sensing has shown that passive sensing features can be used to infer psychological aspects such as mood and stress (person) \cite{Rodriguez2017, LiKamWa2013}, activity levels and types (behavior) \cite{Rabbi2015, SamsungHealth2021}, sociability and social context (person, context) \cite{Berke2011}. In the context of eating behavior, these pillars of data can be mapped to the eight dimensions proposed by Bisogni et al. \cite{bisogni2007dimensions}. Gatica-Perez et al. \cite{gatica2019discovering} showed that mobile sensing features and self-reports could be represented using Bisogni's framework to understand eating routines meaningfully. In essence, this means that smartphone sensing features have shown to be promising in inferring attributes that have shown to be part of the eight dimensions related to eating events. Such relationships have been used in prior ubiquitous computing (ubicomp) studies regarding the eating and drinking behavior \cite{Bae2017, gatica2019discovering, Biel2018, labhart2021ten, Santani2018, meegahapola2021examining, Meegahapola2021OneMoreBite}. Leveraging these relationships, this study seeks to examine whether smartphone sensing could be used to directly infer eating events by taking situational context and behavior sensed via smartphones as proxies for eating events. This study objective is summarized in Figure~\ref{fig:objective}. Hence, we hypothesize that date and time, application usage, screen usage, activity level, and movements are indicative of eating and non-eating.

\subsection{Mobile Food Diaries}\label{subsec:mobile_food_diaries} 
Food diaries are essential to understanding individual and population-level eating practices \cite{jung2020foundations}. While manual food logging using pencil-and-paper based techniques could be useful for self-reflection, mobile food diaries allow logging fine-grained details (weight and size of dishes, variety of dishes) about eating episodes more systematically by searching food types, varieties, and sizes in a database and logging them \cite{jung2020foundations}. Some recent mobile food diaries have also looked into easing the process of logging using speech and photos \cite{luo2021foodscrap}. Going a step further, some studies looked into estimating the calorie intake \cite{okamoto2016automatic} and also calorie deficit by combining information from food diaries and passively detected physical activities \cite{Denning2009BALANCE}. Some popular mHealth apps that provide food journaling functionalities include Samsung Health, Lose It!, MyFitnessPal, EasyDietDiary, and SparkPeople \cite{jung2020foundations, Meegahapola2021OneMoreBite}. However, while food diaries provide many benefits, they also come with a plethora of drawbacks such as tediousness in using and finding the correct food type, difficulty in recording correct dish size, and losing interest in logging over time, among which one of the most common drawbacks is users forgetting to report eating episodes \cite{jung2020foundations, Biel2018}. Hence, to overcome this barrier, researchers have come up with automated eating detection systems using different mobile sensing modalities \cite{Morshed2020, thomaz2015practical, chun2018detecting, zhang2020necksense, Biel2018}. In a nutshell, these systems would detect eating episodes and provide interventions, automatically keep track of events, or remind users to report details about the eating episode on a mobile food diary. The goal of this study, too, is similar. However, the main difference is that this study only use smartphones to make the inference, while previous studies used wearables.

\subsection{Mobile Sensing for Eating Behavior Monitoring}\label{subsec:mobile_sensing_eating}

\begin{table*}[tb]
    \centering
    \caption{Summary of Eating Detection Approaches in Mobile Sensing. LB: Lab-Based and IW: In-The-Wild experiment.}
    
    % \resizebox{\textwidth}{!}{%
    \begin{tabular}{l l l l l l l} 
    
    %\hline 
    
    \rowcolor{Gray}
    \textbf{Study} &
    \textbf{N} &
    \textbf{LB} &
    \textbf{IW} &
    \textbf{Modality (Sensor Location)} & 
    \textbf{Sensors} &
    \textbf{Proxy}
    \\ [0.5ex] 
    \hline
    
    % ----------------------------------------------------------------------
    % Annapurna: An automated smartwatch-based eating detection and food journaling system
    
    % Sen et al. \cite{sen2020annapurna} &
    % 21&
    % \checkmark&
    % \checkmark&
    % Smartwatch (Wrist) &
    % Accelerometer &
    % Hand Movement
    % \\
    
    % ----------------------------------------------------------------------

    \rowcolor{gray!5}
    Morshed et al. \cite{Morshed2020} &
    28& 
    \checkmark&
    \checkmark&
    Smartwatch (Wrist) &
    Accelerometer &
    Hand Movement
    \\
    
    % ----------------------------------------------------------------------
    % Sensing Fork : Eating Behavior Detection Utensil and Mobile Persuasive Game
    
    Kadomura et al. \cite{kadomura2014persuasive} &
    5& 
    &
    \checkmark&
    Sensing Fork (Fork) &
    Single-pixel RGB color sensor&
    Color of the food
    \\
    
    &
    &
    &
    &
    &
    Accelerometer \& Gyroscope &
    Hand Movement
    \\
    
    &
    &
    &
    &
    &
    Three-electrode Conductive Probe &
    Electrical conductivity of food 
    \\
    
    % ----------------------------------------------------------------------
    % iHearFood - Eating Detection Using Commodity Bluetooth Headsets
    
    \rowcolor{gray!5}
    Gao et al. \cite{gao2016ihear} &
    28& 
    \checkmark&
    \checkmark&
    Bluetooth headset (Ear)&
    Microphone &
    Mastication
    \\
    
    \rowcolor{gray!5}
    & 
    & 
    &
    &
    & 
    &
    Deglutition
    \\
    
    % ----------------------------------------------------------------------
    % Eating Episode Detection with Jawbone-Mounted Inertial Sensing
    
    Chun et al. \cite{san2020eating} &
    23&
    \checkmark&
    \checkmark&
    Jawbone-mounted (Jawbone) &
    Accelerometer \& Gyroscope &
    Mastication
    \\
    
    % ----------------------------------------------------------------------
    %Predicting “About-to-Eat” Moments for Just-in-Time Eating Intervention
    
    \rowcolor{gray!5}
    Rahman et al. \cite{rahman2016predicting} &
    8&
    &
    \checkmark&
    Microsoft Band (Wrist) &
    Accelerometer \& Gyroscope &
    Physical Movement
    \\
    
    \rowcolor{gray!5}
    &
    &
    &
    &
    Affectiva Q sensor (Wrist) &
    Electrodermal activity sensor &
    Psychological arousal
    \\
    
    \rowcolor{gray!5}
    &
    &
    &
    &
    Wearable microphone (Neck)&
    Microphone &
    Chewing and swallowing
    \\
    
    \rowcolor{gray!5}
    &
    &
    &
    &
    Smartphone (Smartphone) &
    GPS sensor &
    Location
    \\
    
    % ----------------------------------------------------------------------
    % A Practical Approach for Recognizing Eating Moments with Wrist-Mounted Inertial Sensing
    
    Thomaz et al. \cite{thomaz2015practical} &
    28& 
    \checkmark&
    \checkmark&
    Smartwatch (Wrist) &
    Accelerometer &
    Hand Movement
    \\
    
    % ----------------------------------------------------------------------
    % Auracle: Detecting Eating Episodes with an Ear-mounted Sensor
    
    \rowcolor{gray!5}
    Bi et al. \cite{bi2018auracle} &
    24& 
    \checkmark&
    \checkmark&
    Ear-mounted Sensor (Ear) &
    Microphone &
    Chewing Sound
    \\
    
    % ----------------------------------------------------------------------
    % Assisting Food Journaling with Automatic Eating Detection
    
    Ye et al. \cite{ye2016assisting} &
    7& 
    &
    \checkmark&
    Pebble Smartwatch (Wrist) &
    Accelerometer &
    Hand to mouth eating gestures
    \\
    
    % ----------------------------------------------------------------------
    % MFED: A System for Monitoring Family Eating Dynamics
    
    \rowcolor{gray!5}
    Mondol et al. \cite{mondol2020mfed} &
    74& 
    &
    \checkmark&
    Smartwatch (Wrist) &
    Accelerometer &
    Hand Movement
    \\
    
    % ----------------------------------------------------------------------
    %Toward a Wearable Sensor for Eating Detection
        
    % Bi et al. \cite{bi2017toward} &
    % 20& 
    % &
    % \checkmark&
    % Ear wearable (Ear) &
    % Microphone &
    % Sound 
    % \\
    
    &
    & 
    &
    &
    &
    EMG sensor &
    Motion \& EMG activities
    \\
    
    % ----------------------------------------------------------------------
    % Detecting Eating Episodes by Tracking Jawbone Movements with a Non-Contact Wearable Sensor
    
    \rowcolor{gray!5}
    Chun et al. \cite{chun2018detecting} &
    17& 
    \checkmark&
    \checkmark&
    Necklace (Nect) &
    Proximity sensor &
    Head \& Jawbone movements
    \\
    
    % ----------------------------------------------------------------------
    % Exploring Symmetric and Asymmetric Bimanual Eating Detection with Inertial Sensors on the Wris
        
    Thomaz et al. \cite{thomaz2017exploring} &
    14& 
    \checkmark&
    &
    Wrist-wearable (Wrist) &
    Accelerometer \& Gyroscope &
    Wrist Movements
    \\
    
    % ----------------------------------------------------------------------
    % NeckSense: A Multi-Sensor Necklace for Detecting Eating Activities  in Free-Living Conditions
    
    \rowcolor{gray!5}
    Zhang, et al. \cite{zhang2020necksense} &
    20& 
    &
    \checkmark&
    Necklace (Neck) &
    Accelerometer, Gyroscope, Magnetometer, &
    Head Movement and Chewing
    \\
    
    \rowcolor{gray!5}
    &
    & 
    &
    &
    &
    Proximity, Ambient light &
    \\
    
    % ----------------------------------------------------------------------
    % EarBit: Using Wearable Sensors to Detect Eating Episodes in Unconstrained Environments
    
    Bedri et al. \cite{bedri2017earbit} &
    10& 
    \checkmark&
    \checkmark&
    Ear wearable (Ear) &
    Microphone \& Proximity \& IMU &
    Chewing
    \\
    
    % ----------------------------------------------------------------------
    % Multimodality Sensing for Eating Recognition
    
    \rowcolor{gray!5}
    Merck et al. \cite{merck2016multimodality} &
    6& 
    \checkmark&
    &
    Smartwatch (Wrist) &
    Accelerometer, Gyroscope, Magnetometer &
    Wrist Movements
    \\
    
    \rowcolor{gray!5}
    &
    & 
    &
    &
    Google Glass (Head)&
    Accelerometer, Gyroscope, Magnetometer &
    Head Movements
    \\
    
    \rowcolor{gray!5}
    &
    & 
    &
    &
    Ear wearable (Ear)&
    Microphone &
    Chewing Sounds
    \\
    
    % ----------------------------------------------------------------------
    % Detecting periods of eating during free-living by tracking wrist motion
    
    Dong et al. \cite{dong2013detecting} &
    43& 
    \checkmark&
    &
    Ear wearable (Ear) &
    Microphone \& Proximity \& IMU &
    Chewing
    \\

    % ----------------------------------------------------------------------
    % OUR STUDY

    \rowcolor{red!5}
    Our Study &
    58 & 
    &
    \checkmark&
    Smartphone (Any) &
    Multimodal &
    Behavior and Context
    \\
    \hline 
    \end{tabular}
    \label{tab:related_work}
\end{table*}

Mobile sensing studies for eating behavior monitoring could be segregated into two main categories \cite{Meegahapola2021Survey}: (1) detecting eating events (i.e., time of eating); and (2) characterizing eating events by identifying behavioral and situational context-related routines around eating episodes. Many prior studies using wearable sensing modalities to detect eating episodes fall under the first category. For example, Chun et al. \cite{chun2018detecting} used a necklace-based wearable to detect jawbone and head movements in determining eating episodes. Their technique showed a precision of 95.2\% and a recall of 81.9\% in controlled studies, and precision of 78.2\%, and a recall of 72.5\% in a free-living study. Bedri et al. \cite{bedri2017earbit} studied an ear-worn wearable system called EarBit to detect chewing moments, with 90.1\% and 93\% accuracies in lab settings and outside-lab-settings, respectively. Further, they showed that they could recognize eating episodes ranging from 2-minute snacks to 30-minute meals. Morshed et al. \cite{Morshed2020} used a wristwatch-based eating detection system, obtaining an accuracy of 96.5\% in detecting eating episodes. They also showed how detecting eating episodes using a wearable could trigger momentary ecological assessments (EMA) in the smartphone to capture additional contextual and behavioral information about the eating episode. Thomaz et al. \cite{thomaz2015practical} also showed that it is possible to detect eating episodes using smartwatch-based inertial sensors with F1 scores of 71.3\% and 76.1\% in two experiments done in free-living conditions. Moreover, Rahman et al. \cite{Rahman2016} used a combination of wrist-worn wearables and audio from a mic to predict about-to-eat moments with a recall of 77\%. Further, they showed that personalization could increase the recall up to 81\%. 

Table~\ref{tab:related_work} summarizes the differences between other mobile sensing studies and this study. On a fundamental level, while all other studies primarily used wearables, this study uses commodity smartphones. Further, while many prior studies focused on using sensor data streams from one or two wearable sensing modalities, rich and multimodal sensor data streams coming from smartphones are focused on here. In addition, while most studies attempt to detect eating episodes by sensing actions such as hand movement, chewing, bites, mastication, etc. (hence using them as proxies for eating), we seek to understand eating and non-eating events with behavioral and contextual features captured via smartphone sensors and leverage them to detect eating events.

\subsection{Smartphone Sensing for Eating Behavior Monitoring}\label{subsec:smartphone_sensing_eating}

Smartphone sensing has not been often used to monitor eating behavior. Even the available studies focused on only characterizing eating events. Madan et al. \cite{Madan2010Social} conducted a study to understand the eating behavior of university students in the United States. They concluded that healthy eating behaviors of individuals are related to the health and well-being of others with whom they associate. In another study, Seto et al. \cite{Seto2016} concluded that behavior and context could affect eating patterns. However, these studies did not attempt an inference task using smartphone sensing data. Biel et al. \cite{Biel2018} used smartphone-based contextual sensing to characterize eating events by detecting meal vs. snack events using a time window of two hours around eating episodes to aggregate sensor data. They deployed a mobile application among 122 swiss university students, collecting over 4440 eating episodes. They performed an eating occasion type inference (meal vs. snack) with an accuracy of 84\% with features such as time of the day, time since the last food intake, location, and other sensor data. Meegahapola et al. \cite{Meegahapola2021OneMoreBite} showed that self-reported food consumption levels (eating as usual, overeating, undereating) could be inferred with an accuracy of 83.49\% in a three-class inference task using only smartphone sensing features with a one hour time window around eating episodes. In another study, Meegahapola et al. \cite{Meegahapola2020Alone} showed that the social context of eating events could be inferred with accuracies above 77\% for student populations in two countries. However, all these efforts are towards characterizing eating events, and not detecting them.

The uniqueness of the above studies is that, similar to Bisogni et al. \cite{bisogni2007dimensions}, they consider eating episodes as holistic events that happen amidst different behavioral and situational circumstances, in addition to the main action of eating. Further, they have performed inferences assuming that eating episodes can be detected, including the hour of eating as a feature in inference models. Therefore, the inference can only be made once the eating events are detected. These studies build upon the premise that wearables can detect, and smartphones can characterize eating events. In this study, we consider eating as a holistic event and attempt an eating event detection task that has not been attempted in prior work. Further, since prior work has shown that smartphones fare reasonably well in characterizing eating events, this study complements them well in showing the potential of using only smartphones for detecting eating events. 

\iffalse 
In summary, our study differs from prior work in eating event/episode detection in the following aspects:  

\begin{enumerate}
    \item We use only smartphones for sensing eating events.
    \item Instead of just considering the action of eating (eating episode), we consider eating as a holistic event with interconnected dimensions (including behaviors and contexts around the eating episode) that can be sensed via a smartphone.
    \item We show that eating and non-eating events can be inferred using smartphone sensing and that personalization is crucial in attaining high performance. We also show that feature choice is an important aspect when personalizing. 
    \item While many studies leverage relationships between the sensor data, proxy, and the action of eating, we determine eating and non-eating events using more sparse and noisy multi-modal smartphone sensing data. Hence, this is a more challenging task.
\end{enumerate}
\fi 

Finally, as shown in Figure~\ref{fig:objective}, the methodology was rigorously evaluated using different sensor features, feature selection techniques, model types, and personalization techniques. 
\section{Data, Smartphone Features, and Definition of Eating/non-eating Episodes}\label{sec:mobile_app}

\subsection{Dataset}\label{subsec:dataset}

We used a dataset regarding the eating behavior of young adults from our previous work \cite{Meegahapola2021OneMoreBite}. This dataset consists of self-reports and smartphone sensing data regarding eating and non-eating events from college students (mean age 23.4, 44\% men) in Mexico. A mobile app called i-Log \cite{Zeni2014} was used to collect data from participants. An entry questionnaire was given to participants to collect demographic information. During the deployment, time diaries captured details regarding food intake similar to a mobile food diary and end-of-the-day surveys that captured additional details regarding activity, sleep, and mood. The users were sent notifications three times a day to fill out a food intake questionnaire which asked about the number of eating episodes they had within the last four hours. Then the users were asked to provide data regarding the last food intake which includes how long ago the eating episode had occurred, food categories (meat, fish, bread etc.), social context of eating (alone, with a date, with a group of friends etc.), semantic eating location (home, university, restaurant etc.), concurrent activities (reading, socializing, watching TV etc.), mood and stress level at the time of eating (5-point scale) \cite{Meegahapola2021OneMoreBite}.

Experiments were carried out in two phases. The first phase was from September to October 2019, with the participation of 29 young adults. The second phase was from November to December 2019, where 55 additional young adults took part. Participant inclusion criteria meant that they owned an Android smartphone and did not have eating disorders like bulimia or anorexia. After data cleaning and pre-processing, the final dataset used for this study consists of 12016 self-reports from 58 participants, out of which 1837 are eating events, and 10179 are non-eating events, for an average of 207 total events (eating or non-eating) per participant. 

\begin{table*}

\centering
\caption{Summary of 40 Features Extracted from Smartphone Sensors}
\label{table:allfeatures}
\resizebox{\textwidth}{!}{%
\begin{tabular}{>{\arraybackslash}m{3.1cm} >{\arraybackslash}m{14cm}}

\rowcolor{gray!15}
\textbf{Sensor} & 
{Sensor Description} 
\\

%\rowcolor{gray!10}
{-- Acronym (\# of features)} & 
{Example Features} 
%& \textbf{Group}
\\ 

\hline

\rowcolor{gray!5}
\textbf{Location} &
Using location data, the radius of gyration \cite{Yue2014, Barlacchi2017} associated with the one-hour episode was calculated. It is a commonly used metric in UbiComp research. Moreover, in the calculation, location coordinate values were lowered in precision for location privacy reasons (using only four decimal points). 
\\

%\rowcolor{gray!5}
-- LOC (1) & 
radius\_of\_gyration
\\

%\hline
\rowcolor{gray!5}
\textbf{Accelerometer} & 
For each ten-minute time window of the day, features that represent the mean of all values and mean of absolute values (abs) were generated using accelerometer data for axes x, y, and z separately. Using them, values corresponding to the one-hour eating/non-eating event windows were calculated by taking the mean of six ten-minute time bins. Further, by using the half an hour before and after the T$_{anc}$, more features were generated that correspond to the time before (bef) and after (aft) the eating time \cite{Meegahapola2021OneMoreBite}.   
\\

-- ACC (18) & 
considering the one-hour window: acc$_{x}$, acc$_{y}$, acc$_{z}$, acc$_{x}$abs, acc$_{y}$abs, acc$_{z}$abs
\\ 

& 
before and after T$_{anc}$: acc$_{x}$bef, acc$_{y}$bef, acc$_{z}$bef, acc$_{x}$abs\_bef, acc$_{y}$abs\_bef, acc$_{z}$abs\_bef, acc$_{x}$aft, acc$_{y}$aft, acc$_{z}$aft, acc$_{x}$abs\_aft, acc$_{y}$abs\_aft, acc$_{z}$abs\_aft
\\ 

%\hline 

%\hline 
\rowcolor{gray!5}
\textbf{Application} &
Prior Ubicomp studies have used app usage as a proxy for the behavior of participants \cite{Santani2018}. Similarly, the ten most frequently used apps in the dataset were selected. Then, for each hour of consideration, whether each app was used during the episode was derived.  
\\ 

-- APP (10) &
facebook, whatsapp, instagram, youtube, chrome, spotify, android dialer, youtube music, google quick search box, microsoft launcher
\\

%\hline 
\rowcolor{gray!5}
\textbf{Battery} &
Battery level and charging state have been used as proxies for smartphone usage behavior, which also represents the behavior of study participants \cite{Bae2017, Abdullah2016}. Hence, the average battery level during the one hour was calculated. In addition, whether the phone is charging or not was detected together with a possible source (ac - alternative current, USB, or unknown). 
\\ 

-- BAT (6) &
battery\_level, charging\_true, charging\_false, charging\_ac, charging\_usb, charging\_unknown  
\\

\rowcolor{gray!5}
\textbf{Screen} &
Similar to prior work \cite{Abdullah2016, Bae2017}, screen-on and screen-off events during the one-hour time window were calculated.
\\ 

-- SCR (2)  &
screen\_on, screen\_off
\\

%\hline 
\rowcolor{gray!5}
\textbf{Date and Time} &

The hour of the day and the minute of the day were derived. In addition, another feature captures whether the day is a weekend or not. Prior work has shown that the behavior (e.g., mobility, food consumption, etc.) of people could differ significantly during weekdays and weekends \cite{meegahapola2019buscope}.  
\\ 

-- TIME (3) &
hours\_elapsed, minutes\_elapsed, weekend  
\\

\hline

    \end{tabular}
    }
\end{table*}

\subsection{Collecting Ground Truth and Passive Sensing Data}\label{subsec:mobileapp}
The mobile app was designed to capture retrospective self-reports. Capturing retrospective reports is important because in-situ self-reports might alter the normal behavior of participants during eating episodes, which would add noise to sensed data. In addition, dietary recall techniques are common in eating behavior studies (e.g. 24H dietary recall \cite{liu2011development, castell2015and}). Hence, during three timeslots of the day that are a minimum of four hours apart, the mobile application sent a reminder to participants to report a food intake (note that a far lesser four-hour window was used in this study to capture eating reports compared to a typical 24H recall). The ground truth responses were: \\(a) Case 1: no food intake within the last four hours. \\(b) Case 2: one food intake within the last four hours. \\(c) Case 3: two or more food intakes within the last 4 hours. 

In Case 2 and Case 3, they were asked to report how long ago the last food intake occurred, and the possible answers included 0-30 min, 31-60 min, 60-90 min, 90-120 min, 120-150 min, 150-180 min, 180-210 min, and 210-240 min ago. This report helps to determine an approximate eating time (\textbf{T$_{anc}$}) as an \textit{anchor} for the last eating episode. For example, if a self-report was done at 8.00 pm, and if the eating episode occurred 31-60 minutes ago, the approximate eating time was about 45 minutes ago (mean of 31 and 60), hence \textbf{T$_{anc}$} = 7.15 pm. Further, in Case 2, except for the time window corresponding to that eating episode, the rest of the times correspond to non-eating. Hence, a maximum of two non-eating events from such self-reports were randomly sampled. Furthermore, in Case 1, the last four hours would correspond to a 
{non-eating} period, and a maximum of three time windows were randomly sampled from the last four hours as \textbf{T$_{anc}$} for non-eating events. Moreover, it is worth noting that all eating and non-eating events were chosen such that they are \textit{non-overlapping} when sensor data are aggregated with a time window (described in the next sub section), hence avoiding any biases in the evaluation. While self-reports were captured only three times per day, passive smartphone sensing data were captured throughout the 24 hours. The sensing modalities include the accelerometer (ACC), location (LOC), battery (BAT), screen (SCR), and application usage (APP). A summary of passive smartphone sensing features used in the study is given in Table~\ref{table:allfeatures}.

\subsection{What is an Eating Event?}\label{subsec:eatingepisode}
\begin{wrapfigure}{}{0.5\columnwidth}
\vspace{-0.2 in}
  \begin{center}
    \centering
        \centering
        \includegraphics[width=0.25\textwidth]{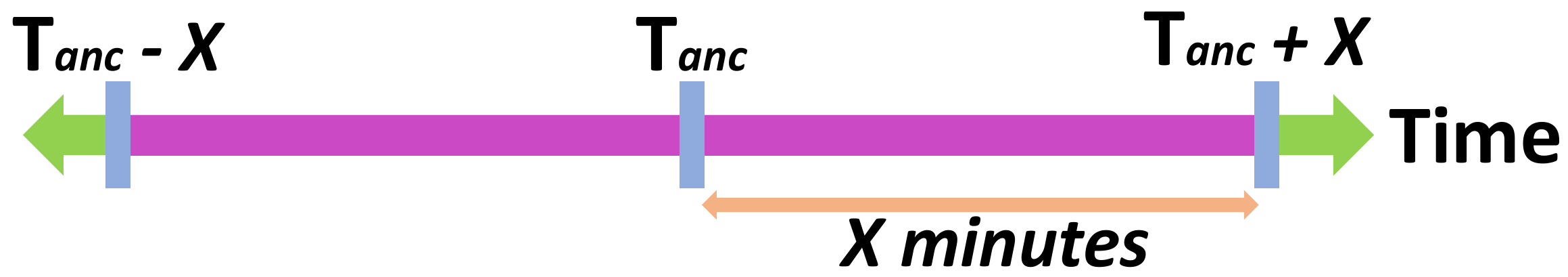}
        \caption{Time Window for Sensor Data Aggregation}
        \label{fig:time_window}
  \end{center}
  \vspace{-0.2 in}
\end{wrapfigure}

A 2X time window around \textbf{T$_{anc}$} was used to aggregate sensor data to match a self-report as given in Table~\ref{table:allfeatures}. According to Figure~\ref{fig:time_window}, 2X time window means that the inference can be done X minutes after the \textbf{T$_{anc}$}. Using a larger time window like one hour (X = 30 minutes) is common for ubicomp studies regarding eating and drinking behavior that considers the context and behavior of participants in addition to the actual eating/drinking episode \cite{Bae2017, Meegahapola2021OneMoreBite, Biel2018, meegahapola2021examining}. This does not mean that the time of inference should be 30 minutes after the end of eating. In reality, it could be much lesser than that because eating episodes can span around 20-30 minutes. Moreover, even though most of the analysis in the next sections focuses on a one-hour time window, the time window can be changed depending on the dataset and application requirements. In Section~\ref{sec:discussion} this is further discussed. 

With the said time window, for the previous example where \textbf{T$_{anc}$} = 7.15pm, sensor data would be aggregated from 6.45pm (\textbf{T$_{anc}$}-30) to 7.45pm (\textbf{T$_{anc}$}+30). In case participants reported that they did not have food during the last four hours, and if the self-report time is \textbf{T$_{sr}$}, \textbf{T$_{anc}$} was chosen carefully to make sure that a half-an-hour window was present on either side of the \textbf{T$_{anc}$}, within the non-eating time window (\textbf{T$_{sr}$-30} >= \textbf{T$_{anc}$} and \textbf{T$_{anc}$} >= \textbf{T$_{sr}$-210}). This is to ensure that there are sufficient sensor data (one-hour) in a non-eating time window to match the self-report. Again, it is worth noting that all the eating and non-eating events in the dataset contain non-overlapping sensor data. Hence, each event is mutually exclusive, and there is no data leakage. Moreover, there is no clear definition for the terms `eating event' or `eating episode' as these terms have been used interchangeably in different studies \cite{bell2020automatic}. Therefore, for clarity, throughout this study, the one-hour time window reported by the participant as they had food during that time is referred as an \textit{\textbf{eating event}}. This event consists of the actual \textit{\textbf{eating episode}} (i.e., the time period of actual eating) and of an extended time period that aims to capture the surrounding behavior and context around the eating episode, similar to prior work \cite{Bae2017, Biel2018, Meegahapola2021OneMoreBite, Meegahapola2020Alone, meegahapola2021examining}. Finally, a one-hour time window in which food is not consumed by participants is referred as a \textit{\textbf{non-eating}} event.

\section{Descriptive and Statistical Analysis of Sensor Data and Eating Events (RQ1)}\label{sec:data_analysis}

\paragraph{Accelerometer Features}\label{subsec:acc_data_distributions}

\begin{figure}[t]
% \begin{center}
    \begin{subfigure}[t]{0.3\columnwidth}
        \centering
        \includegraphics[width=\textwidth]{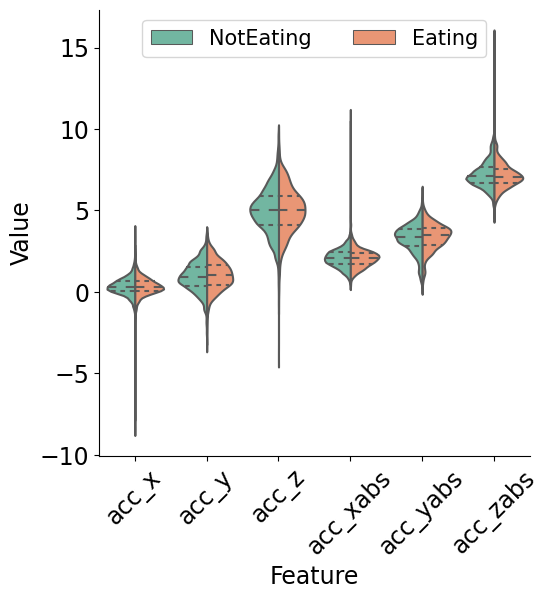}
        \caption{All Users}
        \label{fig:acc_all}
    \end{subfigure}
    \hfill 
    % \begin{subfigure}[t]{0.22\textwidth}
    % \begin{subfigure}[t]{0.3\columnwidth}
    %     \centering
    %     \includegraphics[width=\textwidth]{images/acc_user_5.png}
    %     \caption{User A}
    %     \label{fig:acc_u1}
    % \end{subfigure}
    % \hfill 
    \begin{subfigure}[t]{0.3\columnwidth}
        \centering
        \includegraphics[width=\textwidth]{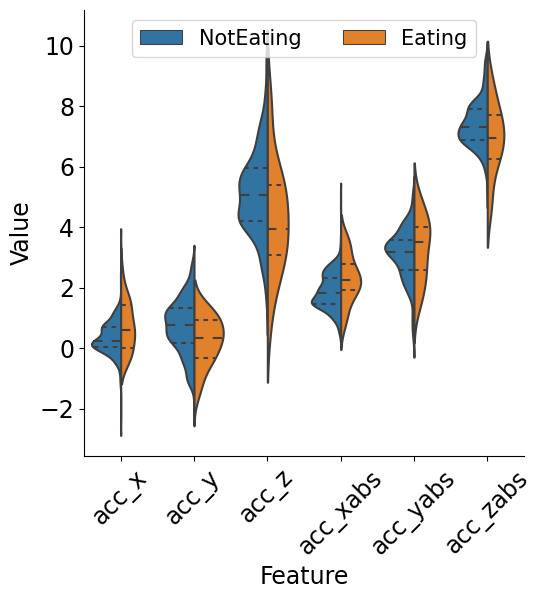}
        \caption{User A}
        \label{fig:acc_u2}
    \end{subfigure}
    \hfill 
    \begin{subfigure}[t]{0.3\columnwidth}
        \centering
        \includegraphics[width=\textwidth]{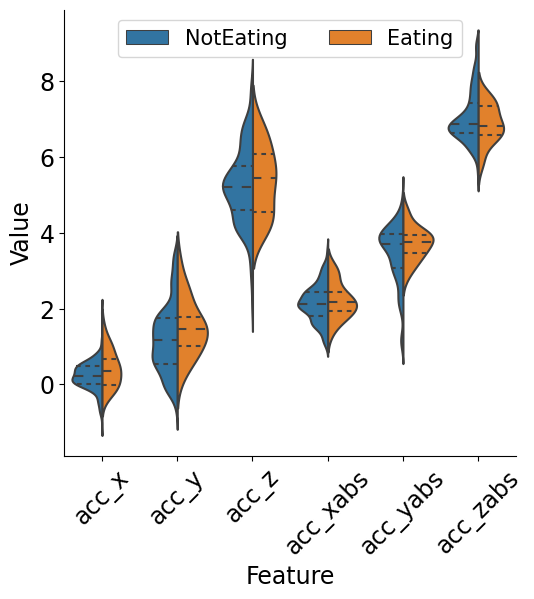}
        \caption{User B} % user with no significant variances
        \label{fig:acc_u3}
    \end{subfigure}
    
    \caption{Violin plots of six selected accelerometer features for all the users and for two randomly selected users}
    \label{fig:acc_data}
% \end{center}
\end{figure}

Figure~\ref{fig:acc_data} shows accelerometer data distributions for all users and two randomly selected users. In Figure~\ref{fig:acc_all}, accelerometer data distributions for eating and non-eating events look similar when all the users are considered in general, with minimal differences between means in all six features. However, for individual users in Figure~\ref{fig:acc_u2} and Figure~\ref{fig:acc_u3}, there are visible differences in accelerometer data distributions. Hence, even though the all-user distribution looks the same for eating and non-eating events, individual-level differences could be leveraged when building inference models by considering within-user differences during eating and non-eating events.

\begin{figure}[t]
% \begin{center}
    % \begin{subfigure}[t]{0.3\textwidth}
    \begin{subfigure}[t]{0.3\columnwidth}
        % \centering
        \includegraphics[width=0.99\textwidth]{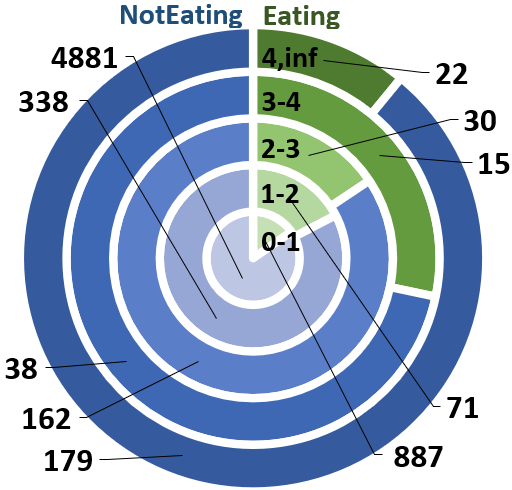}
        \caption{All Users}
        \label{fig:rog_all}
    \end{subfigure}
    \hfill 
    \begin{subfigure}[t]{0.3\columnwidth}
        % \centering
        \includegraphics[width=0.98\textwidth]{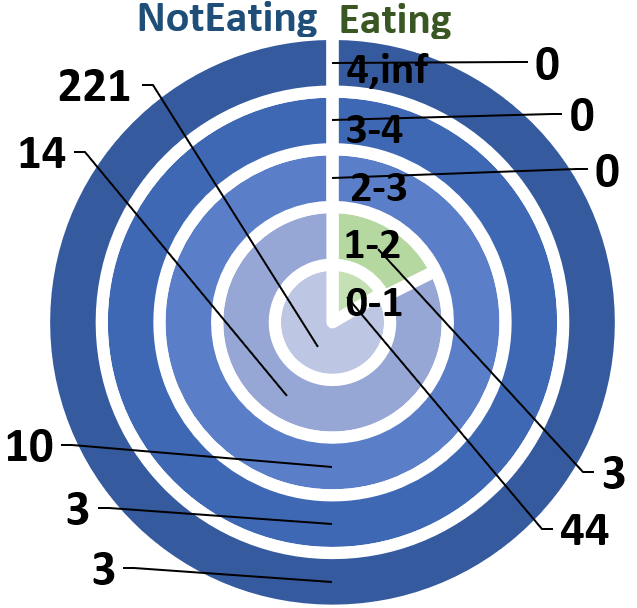}
        \caption{User A}
        \label{fig:rog_u1}
    \end{subfigure}
    \hfill 
    \begin{subfigure}[t]{0.3\columnwidth}
        % \centering
        \includegraphics[width=0.96\textwidth]{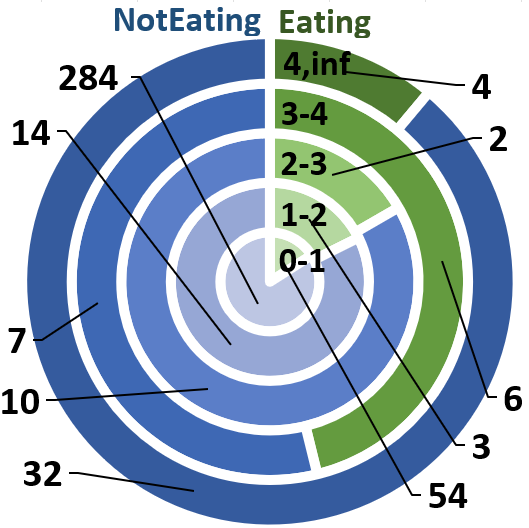}
        \caption{User B}
        \label{fig:rog_u2}
    \end{subfigure}
    \hfill
    \caption{Distribution of eating and non-eating events for different radius of gyration values for all the users and for two randomly selected users}
    \label{fig:eating_ep_vs_rog}
% \end{center}
\vspace{-0.2 in}
\end{figure}

\paragraph{Location Feature}\label{subsec:location_data_distribution}

Figure~\ref{fig:rog_all} shows a distribution of eating and non-eating events for different values of radius of gyration for all users. Although, according to the figure, the ratio of eating to non-eating is almost the same for all radius of gyration values other than between 3-4, a relative increase in eating events can be seen. This suggests that a significant amount of movement during an hour corresponding to eating events. This finding is coherent with prior work that said people might travel a sizable distance at noon in search of food, regardless of the geographical location  \cite{kerr2012predictors, ma2015food}. However, when distributions of two randomly selected users are considered (Figure~\ref{fig:rog_u1} and Figure~\ref{fig:rog_u2}), even though the distribution of \textit{user B} is almost the same as \textit{all-users distribution} in Figure~\ref{fig:rog_all}, \textit{user A} has a different distribution. That user has eating events only from 0 to 2 radius of gyration values, suggesting that while moving \textit{user A} has not eaten. Therefore, these diagrams capture the behavioral diversity of people in terms of movement during a one-hour window and how such movements correspond to eating and non-eating.

\begin{figure}[t]
% \begin{center}
    % \begin{subfigure}[t]{0.245\textwidth}
    \begin{subfigure}[t]{0.33\columnwidth}
        \centering
        \includegraphics[width=\textwidth]{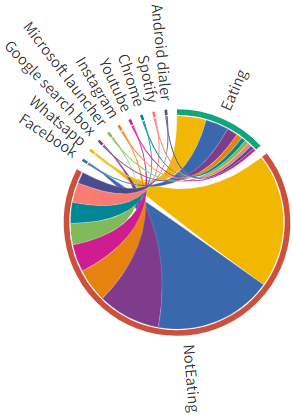}
        \caption{All Users}
        \label{fig:app_all}
    \end{subfigure}
    \hfill 
    \begin{subfigure}[t]{0.31\columnwidth}
        \centering
        \includegraphics[width=0.95\textwidth]{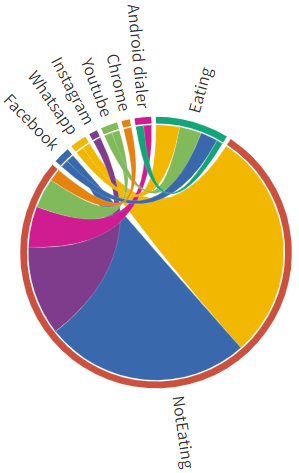}
        \caption{User A}
        \label{fig:app_u1}
    \end{subfigure}
    \hfill 
    \begin{subfigure}[t]{0.33\columnwidth}
        \centering
        \includegraphics[width=\textwidth]{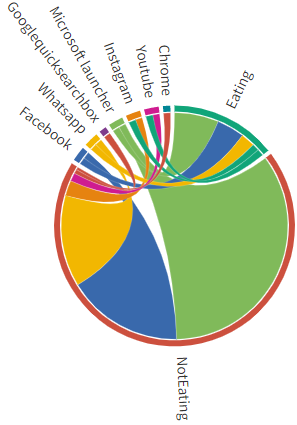}
        \caption{User B}
        \label{fig:app_u2}
    \end{subfigure}
    % \hfill
    % \begin{subfigure}[t]{0.4\columnwidth}
    %     \centering
    %     \includegraphics[width=0.9\textwidth]{images/app_user_3.png}
    %     \caption{User C} % user with no significant variances
    %     \label{fig:app_u3}
    % \end{subfigure}
    \caption{Distribution of app usage for eating and non-eating events for all users and for two randomly selected users}
    \label{fig:eating_ep_vs_apps}
% \end{center}
\vspace{-0.2 in}
\end{figure}

\paragraph{Application Usage Features}\label{subsec:app_usage_distributions}
Figure~\ref{fig:app_all} provides the app usage distribution for eating and non-eating events for all the users. Results suggest that the proportion of app usage for a particular app could differ between eating and non-eating. For example, more people used Spotify than Youtube during eating events, whereas there was more Youtube usage compared to Spotify during non-eating periods. While these population-level statistics look interesting, individual-level app distributions for two random users in Figure~\ref{fig:app_u1} and Figure~\ref{fig:app_u2} suggest that individual-level app usage could alter from the population level. For example, \textit{user A} has not used Instagram and Chrome while eating. They have used Youtube more than Facebook while eating whereas used Facebook more than Youtube when not eating. \textit{User B} has also used Chrome only while not eating. These individual app usage differences are in line with prior work in mobile sensing that has shown that app usage behavior could be used to identify users \cite{do2010their}. Furthermore, app usage behavior has shown good performance for other eating and drinking behavior-related tasks \cite{Santani2018}. In addition, prior work has proposed that app usage can be used to gain an understanding of user context \cite{bohmer2011falling}. Hence, when combined with findings in prior work, the descriptive results indicate why person-level models could be useful over population-level models to capture these fine-grained behavioral differences among people, especially when considering app usage behavior.

In summary, descriptive analysis of the dataset shows that features from passive sensing modalities could indicate eating and non-eating events, especially when combined with the time of the day. Further, results for some modalities such as APP, LOC, and ACC suggest that only considering data from a single user could capture individual-level behavioral differences with regard to eating and non-eating events, that would otherwise be not noticeable when considering population-level behaviors. In the next section, these feature-level relationships are examined and quantified in more detail using various statistical techniques.

\paragraph{Statistical Significance of Features} Table~\ref{tab:statistical_significance} shows the top ten features that help discriminate eating and non-eating, when sorted in the descending order according to the t-statistic \cite{Kim2015}. In addition, the p-values \cite{Greenland2016} are also given. Further, since prior work has highlighted the lack of informativeness in p-values \cite{Lee2016}, we also provide Cohen's-d (effect size) \cite{Rice2005} with 95\% confidence interval (CI) \cite{Lakens2013} to quantify the statistical significance of features. The rule of thumb for interpreting Cohen's-d values is as follows: 0.2 = small effect size, 0.5 = medium effect size, and 0.8 = large effect size. In addition, if the confidence interval does not overlap with zero, the difference between the two groups could be considered significant \cite{Lee2016}. Hence, for the two-class problem of detecting eating and non-eating, this analysis aims to examine features that help discern between the two classes. 

The top feature by t-statistics is acc\_y. Mean values for eating (1.04) and non-eating (0.92) for acc\_y, show that the mobile phone activity around the y-axis is higher when eating than during non-eating events. This observation is moderately supported by the t-statistic (5.02) and effect size (0.12) values. Moreover, similar behavior can be observed in the next top five features, all of which had t-statistics above three. However, all of them had below small effect sizes. The WhatsApp app usage and the feature that captures whether the day is a weekend or not too had t-statistic values above two. Furthermore, all features included in Table~\ref{tab:statistical_significance} had 95\% confidence intervals not overlapping zero. These results suggest that smartphone sensing features from modalities such as ACC, TIME, and APP could be useful to discern between eating and non-eating events, albeit with limited statistical significance when considering the full participant cohort.

\begin{table*}[t]
        \small
        \centering
        \caption{t-statistic, p-value (p-value$\leq$10$^{-4}$:****; p-value$\leq$10$^{-3}$:***; p-value$\leq$10$^{-2}$:**), and Cohen's-d with 95\% confidence interval calculated using 12016 datapoints. Top 10 features are shown in the decreasing order of t-statistic.}
        \resizebox{\textwidth}{!}{%
        \begin{tabular}{l l l l l l l l l l l l l l l l l}

        \rowcolor{gray!15}
        \textbf{Feature} &
        \multicolumn{2}{c}{\textbf{eating}} &
        \multicolumn{2}{c}{\textbf{non-eating}} &
        \textbf{t-statistic} &
        \textbf{p} &
        \textbf{cohen's-d} &
        &
        \textbf{Feature} &
        \multicolumn{2}{c}{\textbf{eating}} &
        \multicolumn{2}{c}{\textbf{non-eating}} &
        \textbf{t-statistic} &
        \textbf{p} &
        \textbf{cohen's-d}
        \\

        \rowcolor{gray!15}
        &
        % &
        \textbf{mean} &
        \textbf{std} &
        % &
        \textbf{mean} &
        \textbf{std} &
        &
        &
       \textbf{[95\% CI]} &
        &
        &
        \textbf{mean} &
        \textbf{std} &
        % &
        \textbf{mean} &
        \textbf{std} &
        &
        &
        \textbf{[95\% CI]}
        \\

        \hline 

        acc$_{y}$ & % feature name
        % ACC& % feature group
        1.04 & % eating mean
        0.89 & % eating std
        % & % empty column
        0.92 & % non-eating mean
        0.92 & % non-eating std
        (+) 5.0199 & % t-statistic
        ****& % p-value
        0.1276 [0.08, 0.18] &% cohen's-d with 95% confidence interval
        
        & % empty column
        
        acc$_{y}$abs\_aft & % feature name 
        % ACC & % feature group
        3.39 & % eating mean
        0.92 & % eating std
        % & % empty column
        3.30 & % non-eating mean
        0.98 & % non-eating std
        (+) 3.4838 & % t-statistic
        *** & % p-value
        0.0897 [0.04, 0.14] % cohen's-d with 95% confidence interval
        \\

        \rowcolor{gray!5}
        acc$_{y}$abs\_bef & % feature name 
        % ACC & % feature group
        3.37 & % eating mean
        0.93 & % eating std
        % & % empty column
        3.24 & % non-eating mean
        1.01 & % non-eating std
        (+) 4.9662 & % t-statistic
        **** & % p-value
        0.1292 [0.08, 0.18] & % cohen's-d with 95% confidence interval
        
        & % empty column
        
        acc$_{z}$abs\_bef & % feature name
        % ACC & % feature group
        7.15 & % eating mean
        0.83 & % eating std
        % & % empty column
        7.23 & % non-eating mean
        0.95 & % non-eating std
        (-) 3.4267 & % t-statistic
        *** & % p-value
        0.0904 [0.04, 0.14] % cohen's-d with 95% confidence interval
        \\

        acc$_{y}$abs & % feature name
        % ACC & % feature group
        3.38 & % eating mean
        0.83 & % eating std
        % & % empty column
        3.27 & % non-eating mean
        0.91 & % non-eating std
        (+) 4.5727 & % t-statistic
        **** & % p-value
        0.1189 [0.07, 0.17] & % cohen's-d with 95% confidence interval
        
        & % empty column
        
        whatsapp & % feature name
        % APP & % feature group
        0.13 & % eating mean
        0.34 & % eating std
        % & % empty column
        0.11 & % non-eating mean
        0.31 & % non-eating std
        (+) 2.8487 & % t-statistic
        ** & % p-value
        0.0698 [0.02, 0.12] % cohen's-d with 95% confidence interval
        \\

        \rowcolor{gray!5}
        acc$_{y}$aft & % feature name
        % ACC & % feature group
        1.06 & % eating mean
        1.06 & % eating std
        % & % empty column
        0.94 & % non-eating mean
        1.06 & % non-eating std
        (+) 4.5035 & % t-statistic
        **** & % p-value
        0.1135 [0.06, 0.16] &% cohen's-d with 95% confidence interval
        
        & % empty column
        
        weekend & % feature name
        % TIME & % feature group
        0.03 & % eating mean
        0.04 & % eating std
        % & % empty column
        0.30 & % non-eating mean
        0.46 & % non-eating std
        (-) 2.7505 & % t-statistic
        ** & % p-value
        0.0703 [0.02, 0.12] % cohen's-d with 95% confidence interval
        \\

        acc$_{y}$bef & % feature name
        % ACC & % feature group
        1.01 & % eating mean
        0.99 & % eating std
        % & % empty column
        0.90 & % non-eating mean
        1.03 & % non-eating std
        (+) 4.3225 & % t-statistic
        **** & % p-value
        0.1103 [0.06, 0.16] & % cohen's-d with 95% confidence interval
        
        & % empty column
        
        acc$_{z}$abs & % feature name
        % ACC & % feature group
        7.15 & % eating mean
        0.75 & % eating std
        % & % empty column
        7.21 & % non-eating mean
        0.08 & % non-eating std
        (-) 2.7364 & % t-statistic
        ** & % p-value
        0.0709 [0.02, 0.12] % cohen's-d with 95% confidence interval
        \\
        
        % \rowcolor{gray!5}
        % acc y abs after & % feature name
        % % ACC & % feature group
        % 3.38904 & % eating mean
        % 0.92483 & % eating std
        % & % empty column
        % 3.30371 & % non-eating mean
        % 0.97782 & % non-eating std
        % (+) 3.48380 & % t-statistic
        % *** & % p-value
        % 0.08966 [0.04015, 0.13916] % cohen's-d with 95% confidence interval
        % \\
        
        % acc z abs before & % feature name
        % % ACC & % feature group
        % 7.14562 & % eating mean
        % 0.82863 & % eating std
        % & % empty column
        % 7.22599 & % non-eating mean
        % 0.94573 & % non-eating std
        % (-) 3.42670 & % t-statistic
        % *** & % p-value
        % 0.09040 [0.04089, 0.13990] % cohen's-d with 95% confidence interval
        % \\
        
        % \rowcolor{gray!5}
        % whatsapp & % feature name
        % % APP & % feature group
        % 0.12919 & % eating mean
        % 0.33550 & % eating std
        % & % empty column
        % 0.10668 & % non-eating mean
        % 0.30872 & % non-eating std
        % (+) 2.84873 & % t-statistic
        % ** & % p-value
        % 0.06983 [0.02033, 0.11933] % cohen's-d with 95% confidence interval
        % \\
        
        % weekend & % feature name
        % % TIME & % feature group
        % 0.026865 & % eating mean
        % 0.044338 & % eating std
        % & % empty column
        % 0.30033 & % non-eating mean
        % 0.45842 & % non-eating std
        % (-) 2.75045 & % t-statistic
        % ** & % p-value
        % 0.07025 [0.02075, 0.11975] % cohen's-d with 95% confidence interval
        % \\
        
        % \rowcolor{gray!5}
        % acc z abs & % feature name
        % % ACC & % feature group
        % 7.15108 & % eating mean
        % 0.74810 & % eating std
        % & % empty column
        % 7.20638 & % non-eating mean
        % 0.080929 & % non-eating std
        % (-) 2.73635 & % t-statistic
        % ** & % p-value
        % 0.07096 [0.02146, 0.12046] % cohen's-d with 95% confidence interval
        % \\ 
        
        \hline 
        
        \end{tabular}}
        \label{tab:statistical_significance}

\end{table*}

\section{Detecting Eating Events and Important Features (RQ2)}\label{sec:inference}

\subsection{{Two-Class Eating Event Inference}}\label{subsec:inference}

\begin{figure}[t]
% \begin{center}
    \begin{minipage}[t]{\columnwidth}
        \centering
        \includegraphics[width=\textwidth]{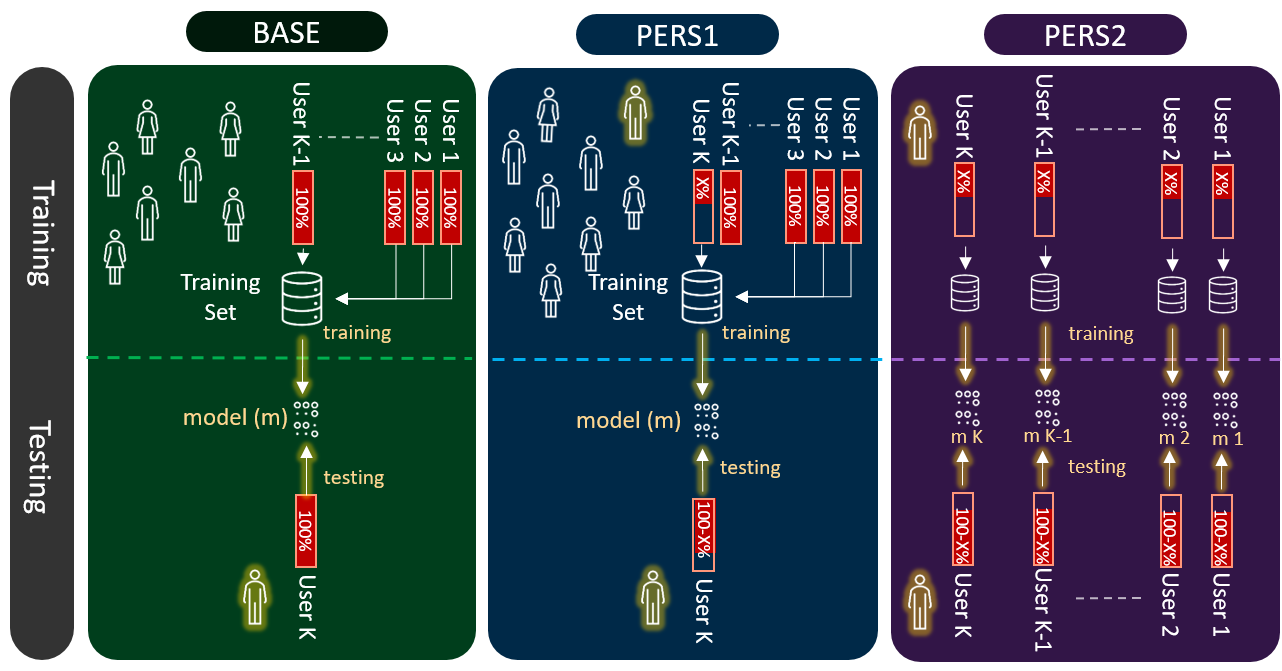}
        \caption{Three Phases of the Study}
        \label{fig:phases}
    \end{minipage}
% \end{center}
\vspace{-0.2 in}
\end{figure}

In this section, we used different feature group combinations to infer eating vs. non-eating events using smartphone sensing data. Scikitlearn framework \cite{scikit-learn} and python is used to conduct experiments in three phases using different model types: (1) Random Forest (RF) \cite{Cutler2011}, (2) Naive Bayes (NB) \cite{Rish2001}, (3) Gradient Boosting (GB) \cite{Natekin2013}, and (4) AdaBoost (AB) \cite{Schapire2013}. These models were chosen by considering the tabular nature of the dataset, interpretability of results (e.g., getting feature importance values), and the small size of the dataset. Further, similar to recent ubicomp work \cite{Bae2018}, Synthetic Minority Over-sampling Technique (SMOTE) \cite{Chawla2002} was used to prepare training sets for each inference task. In addition, F1-Score and AUROC score, both with macro averaging is reported. This would give equal emphasize to both classes, hence indicating whether both eating and non-eating classes are classified well. Moreover, the three phases of experiments are described below (named as subject dependent, subject independent, and hybrid in line with prior work \cite{ferrari2020personalization}) and summarized in Figure~\ref{fig:phases}.

\textbf{Subject Independent (BASE):} These models are also called population-level models. The leave-one-out cross-validation strategy that is commonly used in mobile sensing research \cite{LiKamWa2013, Meegahapola2021OneMoreBite} is used in this phase. The objective is to train on a set of users and test on a user not seen on the training set. Hence, this is the base accuracy because, from a mobile sensing standpoint, this corresponds to a situation where a new user is starting to use a mHealth app, and the server does not have any data from the user. Therefore, the machine learning model used on the user's app is a general model trained with data from other users. This is the accuracy that can be expected for a new user without any personalization.  

\textbf{Hybrid (PERS1):} This corresponds to a situation where the server has some data from a new user to include in training a partially personalized model. However, the server does not have enough data from that user to train a separate, fully personalized model. From a mobile sensing standpoint, this corresponds to a situation where users have used a mHealth app for some time, hence generating some data for model training. Then, users use the mHealth app that contains the partially personalized model. The generated model is partially personalized because user data has been used in training the model, and the same user's data would be used in testing. When conducting experiments, it was ensured that each training split contained data from other subjects (similar to leave-one-out cross-validation) and 70\% data from the target user, and the rest of the data points of that user were used to test the model.

\textbf{Subject Dependent (PERS2):} These models are also called user-level models. The server has enough data to train entirely personal models for each user. Hence, testing is done with a model that is trained with the same user's data. This corresponds to a fully personalized scenario from a mobile sensing standpoint \cite{LiKamWa2013} where users have used a mHealth app for some time and have produced enough data for the server to generate a fully personalized model. In this case, different users would have different models. Hence the approach was evaluated by training the model using 70\% data from each user and testing with the rest of the data of that user, and finally averaging the results of all the users. These models are also called user-level models.

We conducted all three experiments using several feature group combinations. The first task was to understand whether single feature modalities could be used for the inference task. This is important because prior work has shown that having multiple models that can make the same inference could be useful for robust mobile sensing systems in situations of sensor failure \cite{Santani2018}. For instance, a college student might turn off the location sensor during some hours because that sensor could drain the battery faster. In such situations, having different inference models that use other data sources to make the exact inference is crucial. In addition, the feature group TIME would indicate whether time alone could be a good predictor for eating and non-eating events. Hence, the experiments were conducted for feature groups LOC, SCR, TIME, BAT, APP, and ACC separately.

\begin{table}[t]
        \small
        \centering
        \caption{Averaged F1-score (F1) and AUROC of 58 users, calculated using four different models, for the BASE of eating event detection task. For ALL-PCA, the number in square brackets (e.g. [12], [10], etc.) indicate the number of principal components used for the inference. For ALL-FS, the notation in square brackets (e.g. [F3], [F7], etc.) indicate the name of the feature group. More details about feature groups, including the list of features in each group are given in Appendix.}
        \resizebox{0.5\textwidth}{!}{%
        \begin{threeparttable}
        \begin{tabular}{l l l l l}

        \rowcolor{gray!15}
        \textbf{Feature Group}& 
        \textbf{RF} &
        \textbf{NB} &
        \textbf{GB} &
        % \textbf{XGB}&
        \textbf{AB} 
        \\
        
        \rowcolor{gray!15}
        \textbf{(\# of Features)}& 
        \textbf{F1, AUROC} &
        \textbf{F1, AUROC} &
        \textbf{F1, AUROC} &
        % \textbf{A\%, F\%, AR, CK} &
        \textbf{F1, AUROC}
        \\

        \hline 
        
        \rowcolor{gray!5}
        Majority Class &
        0.00, 0.50 & % RF
        0.00, 0.50 & % Naive Bayes
        0.00, 0.50 & % Gradient Boost
        % 19.2, 0.0, 0.50, 0.00 & % XG Boost
        0.00, 0.50 % AdaBoost
        \\

        LOC (1) &
        0.72, 0.50 & % RF
        0.72, 0.48 & % Naive Bayes
        0.72, 0.51 & % Gradient Boost,0
        %  & % XG Boost
        0.72, 0.52 % AdaBoost
        \\
        
        \rowcolor{gray!5}
        SCR (2) &
        73.0, 0.52 & % RF
        72.7, 0.54 & % Naive Bayes
        72.7, 0.53 & % Gradient Boost
        %  & % XG Boost
        72.7, 0.54 % AdaBoost
        \\
        
        TIME (3)&
        0.72, 0.51 & % RF
        0.72, 0.54 & % Naive Bayes
        0.72, 0.52 & % Gradient Boost
        %  & % XG Boost
        0.72, 0.53 % AdaBoost
        \\
        
        \rowcolor{gray!5}
        BAT (6) &
        0.71, 0.51 & % RF
        0.72, 0.50 & % Naive Bayes
        0.72, 0.52 & % Gradient Boost
        %  & % XG Boost
        0.72, 0.53 % AdaBoost
        \\

        APP (10) &
        0.72, 0.51 & % RF
        0.70, 0.51 & % Naive Bayes
        0.72, 0.51 & % Gradient Boost
        %  & % XG Boost
        0.72, 0.51 % AdaBoost
        \\
        
        \rowcolor{gray!5}
        ACC (18) &
        0.72, 0.55 & % RF
        0.72, 0.55 & % Naive Bayes
        0.72, 0.54 & % Gradient Boost
        %  & % XG Boost
        0.72, 0.52 % AdaBoost
        \\

        ALL (40) &
        0.74, 0.65 & % RF
        0.70, 0.56 & % Naive Bayes
        0.72, 0.59 & % Gradient Boost
        %  & % XG Boost
        0.72, 0.56 % AdaBoost
        \\
        
        \rowcolor{gray!5}
        ALL-PCA &
        0.73, 0.53 [c=5] & % RF
        0.72, 0.53 [c=1] & % Naive Bayes
        0.72, 0.52 [c=4] & % Gradient Boost
        %  & % XG Boost
        0.72, 0.52 [c=4] % AdaBoost
        \\
        
        ALL-FS &
        0.74, 0.65 [F1] & % RF
        0.71, 0.51 [F7] & % Naive Bayes
        0.72, 0.58 [F3] & % Gradient Boost
        %  & % XG Boost
        0.72, 0.50 [F7] % AdaBoost
        \\
        
        \hline 
        
        \end{tabular}
        \iffalse
        \begin{tablenotes}
           \item[1] {\scriptsize screen\_on\_count, screen\_off\_count, facebook, whatsapp, googlequicksearchbox, microsoft\_launcher, instagram, youtube, chrome, spotify, android\_dialer, youtube\_music, battery\_level, charging\_true\_count, charging\_false\_count, charging\_ac, charging\_usb, charging\_unknown minutes\_elapsed, hours\_elapsed, weekend, acc\_x\_bef, acc\_y\_bef, acc\_z\_bef, acc\_x\_aft, acc\_y\_aft, acc\_z\_aft, acc\_yabs, acc\_zabs, acc\_xabs\_bef, acc\_yabs\_bef, acc\_xabs\_aft, acc\_yabs\_aft, acc\_zabs\_aft, radius\_of\_gyration}
           \item[2] {\scriptsize screen\_on\_count, screen\_off\_count, facebook, whatsapp, googlequicksearchbox, microsoft\_launcher, instagram, youtube, chrome, spotify, android\_dialer, youtube\_music, charging\_true\_count, charging\_false\_count, charging\_ac, charging\_usb, charging\_unknown, minutes\_elapsed, hours\_elapsed, weekend, acc\_z\_bef, acc\_x\_aft, acc\_z\_aft, acc\_yabs, radius\_of\_gyration}
           \item[3] {\scriptsize  googlequicksearchbox, microsoft\_launcher, instagram, youtube, charging\_false\_count}
        \end{tablenotes}
        \fi 
        \end{threeparttable}
        }
        \label{tab:inference_results}
\vspace{-0.1 in}
\end{table}

There are three more feature groups where all features are used. First, the ALL feature group considers all the features available for the inference task. Then ALL-PCA used principal component analysis \cite{wold1987principal} to obtain the optimum number of principal components to get the best accuracy using all features. Then, ALL-FS used a sequential forward feature selection algorithm \cite{marcano2010feature} to select the best set of features for a given inference task. For BASE and PERS1, feature group after feature selection is common for all users because there is only one model for all users. However, since there is one model for each user in PERS2, different feature groups are used to train models after feature selection.

\begin{table}[t]
        % \small
        \centering
        \caption{Averaged F1-score (F1) and AUROC calculated using random forest classifiers, for BASE, PERS1, and PERS2 of the eating event detection task of 58 users. For ALL-FS, the notation in square brackets (i.e. [F1]) indicate the name of the feature group. More details about the feature group, including the list of features in the group is given in Appendix.}
        \resizebox{0.8\columnwidth}{!}{%
        \begin{tabular}{l l l l}

        \rowcolor{gray!15}
        \textbf{Feature Group}& 
        \textbf{BASE} &
        \textbf{PERS1} &
        \textbf{PERS2} 
        \\
        \rowcolor{gray!15}
        \textbf{(\# of Features)}& 
        \textbf{F1, AUROC} &
        \textbf{F1, AUROC} &
        \textbf{F1, AUROC} 
        \\

        \hline 
        Majority Class &
        0.00, 0.50 &
        0.00, 0.50 & 
        0.00, 0.50 
        \\

        LOC (1) &
        0.72, 0.50 &
        0.72, 0.52 & 
        0.74, 0.59
        \\
        
        \rowcolor{gray!5}
        SCR (2) &
        0.73, 0.52 &
        0.72, 0.52 & 
        0.73, 0.57
        \\
        
        TIME (3) &
        0.72, 0.51 &
        0.71, 0.52 & 
        0.72, 0.52
        \\
        
        \rowcolor{gray!5}
        BAT (6) &
        0.71, 0.51 &
        0.72, 0.56 & 
        0.73, 0.60
        \\

        APP (10) &
        0.72, 0.51 &
        0.72, 0.51 & 
        0.71, 0.51
        \\
        
        \rowcolor{gray!5}
        ACC (18) &
        0.72, 0.55 &
        0.75, 0.66 & 
        0.79, 0.73
        \\
        
        ALL (40) &
        0.73, 0.65 &
        0.81, 0.81 & 
        0.80, 0.73
        \\
        
        \rowcolor{gray!5}
        ALL-PCA &
        0.73, 0.53 [c=5] &
        0.79, 0.72 [c=4] &
        0.81, 0.76
        \\
        
        % feature with highest F1-socre
        ALL-FS &
        \textbf{0.74, 0.65} [F1] &
        \textbf{0.81, 0.81} [F1] & 
        \textbf{0.85, 0.81}
        \\
        \hline 
        
        \end{tabular}}
        \label{tab:inference_results_pers}

\vspace{-0.2 in}
\end{table}

Table~\ref{tab:inference_results} and Table~\ref{tab:inference_results_pers} summarize the results of experiments. First, in Table~\ref{tab:inference_results}, BASE results are shown for all inference models. As per the results, the best-performing models are different for different feature groups. The highest F1-score of 0.74 and AUROC of 0.65 for the BASE came from the RF when using the ALL feature group. These results suggest that BASE scores for the inference tasks are moderate. Then, personalized results (PERS1 and PERS2) are included in Table~\ref{tab:inference_results_pers}. Considering space limitations, this table only contains results from RF because they provided the best performance for PERS1 and PERS2 in most cases. There is a bump in AUROC scores for PERS1 compared to BASE in most cases. In addition, for ALL, ALL-PCA and ALL-FS, F1 scores increased by over 6\%. ALL-FS with F1 feature set provided the best F1 score of 0.81. Finally, the best F1 score out of all inference tasks (0.85) came from PERS2 with the ALL-FS feature group. Similar patterns could be seen with AUROC scores. Since PERS2 averages results from 58 different users, the model for each user had different feature sets after feature selection that provided the best performance. Section~\ref{subsec:feature_importance} discusses more details about these feature sets. This suggests that while subject-dependent models could help detect eating events with reasonably high performance, depending on the user, it is better to select the best set of features using a feature selection technique. In addition, hybrid models perform fairly well compared to subject-dependent models for most feature groups.

\subsection{{Feature Importance for Eating event Detection (RQ2)}}\label{subsec:feature_importance}

\begin{figure*}[t]
\begin{center}

    \begin{subfigure}[t]{0.49\textwidth}
        \centering
        \includegraphics[width=\textwidth]{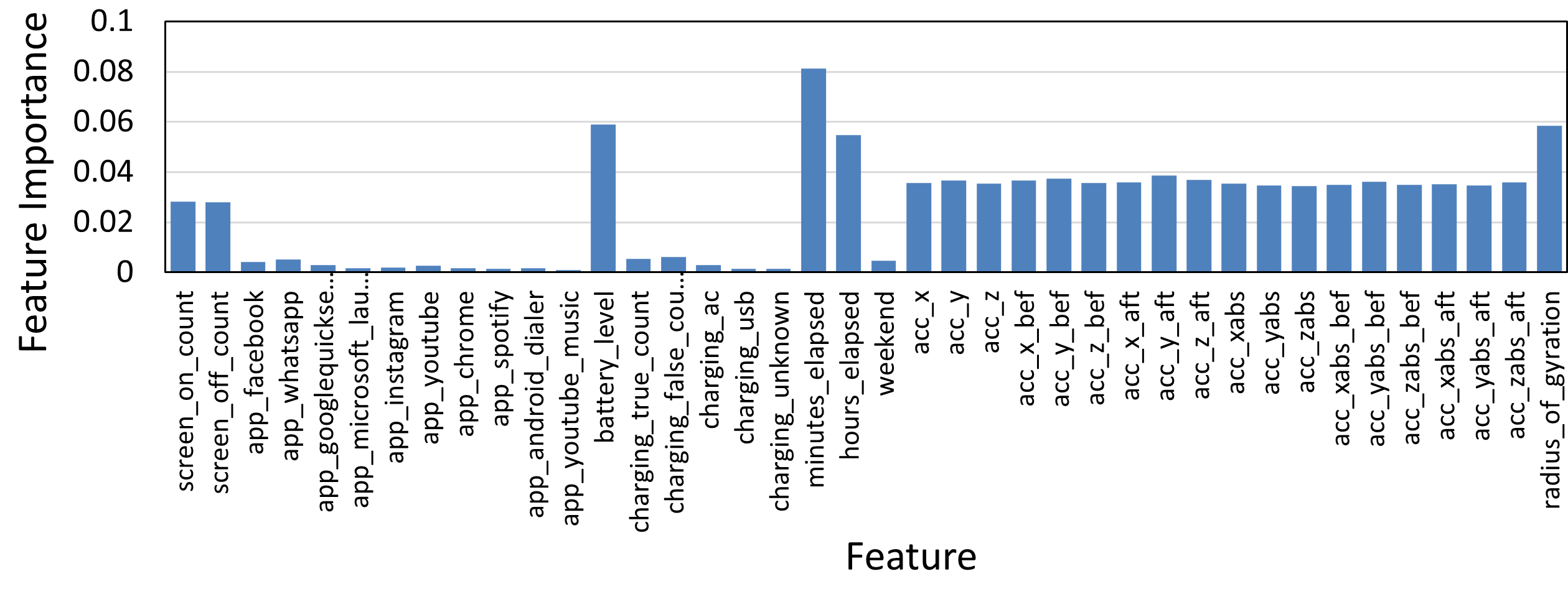}
        \caption{BASE}
        \label{fig:fi_base}
    \end{subfigure}
    \hfill 
    \begin{subfigure}[t]{0.49\textwidth}
        \centering
        \includegraphics[width=\textwidth]{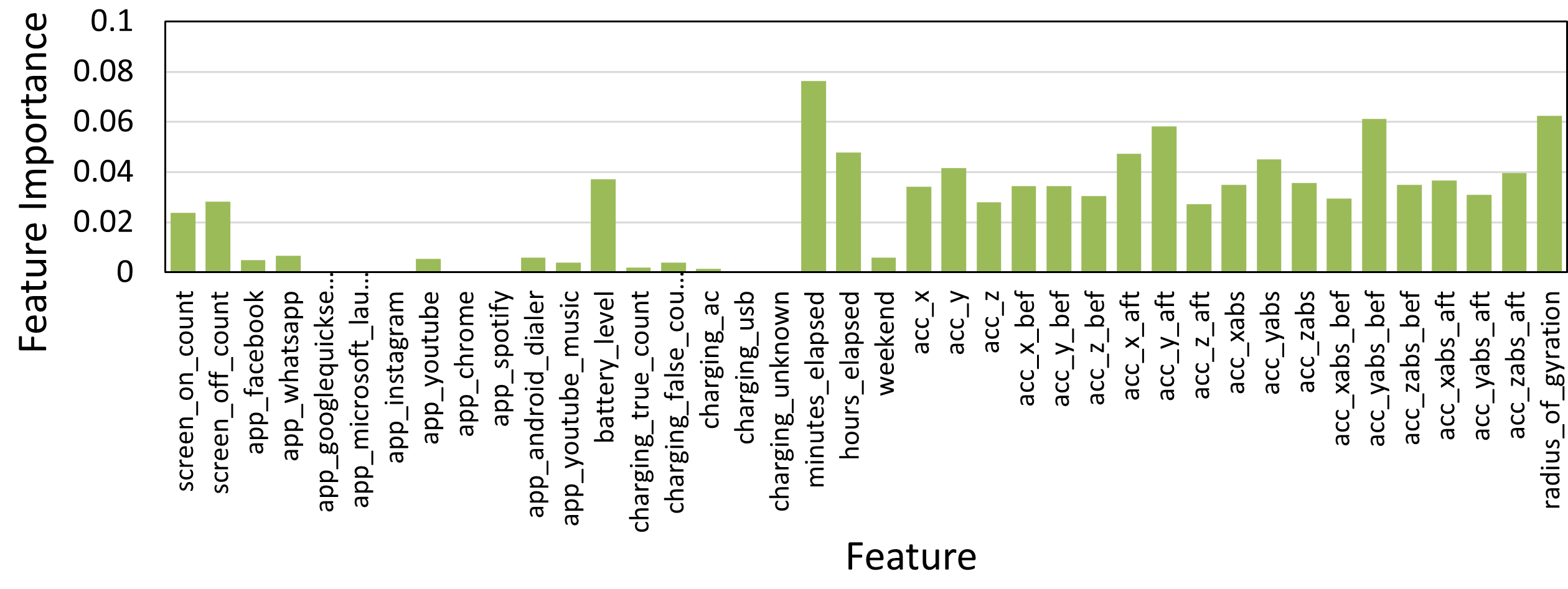}
        \caption{User A}
        \label{fig:fi_user1}
    \end{subfigure}
    \hfill 
    \begin{subfigure}[t]{0.49\textwidth}
        \centering
        \includegraphics[width=\textwidth]{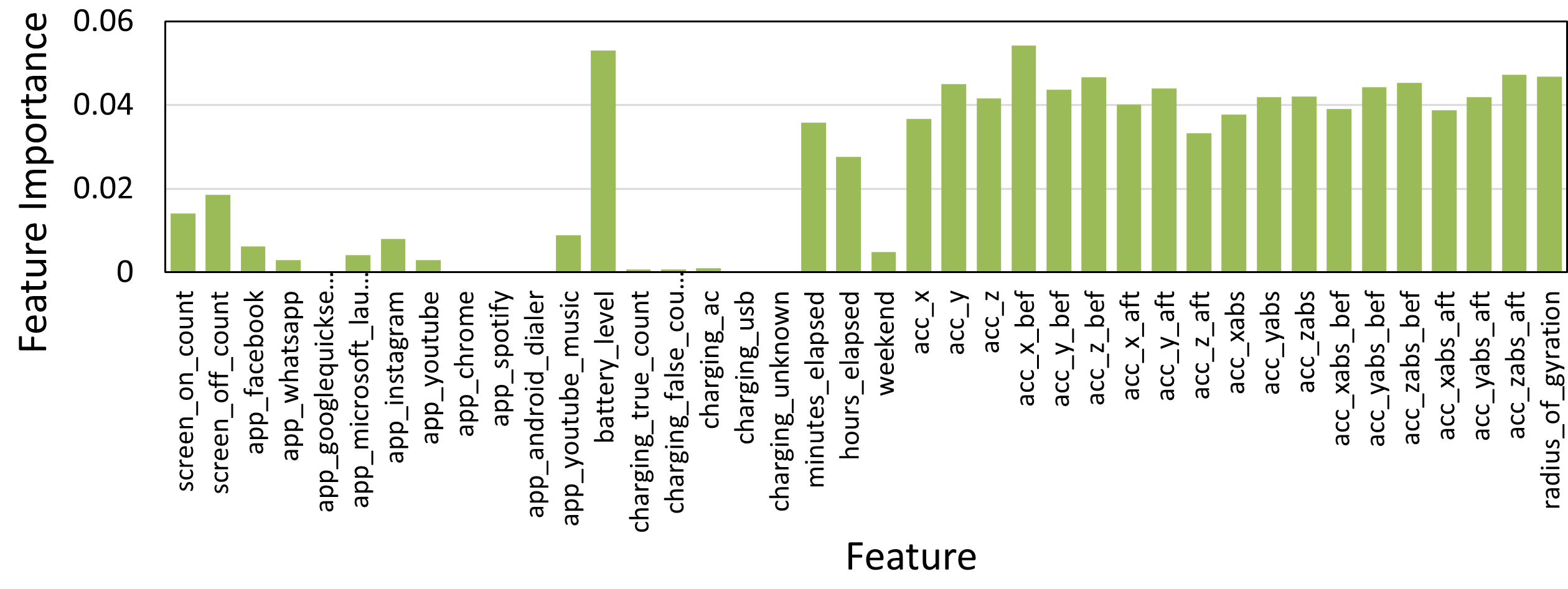}
        \caption{User B}
        \label{fig:fi_user2}
    \end{subfigure}
    \hfill 
    \begin{subfigure}[t]{0.49\textwidth}
        \centering
        \includegraphics[width=\textwidth]{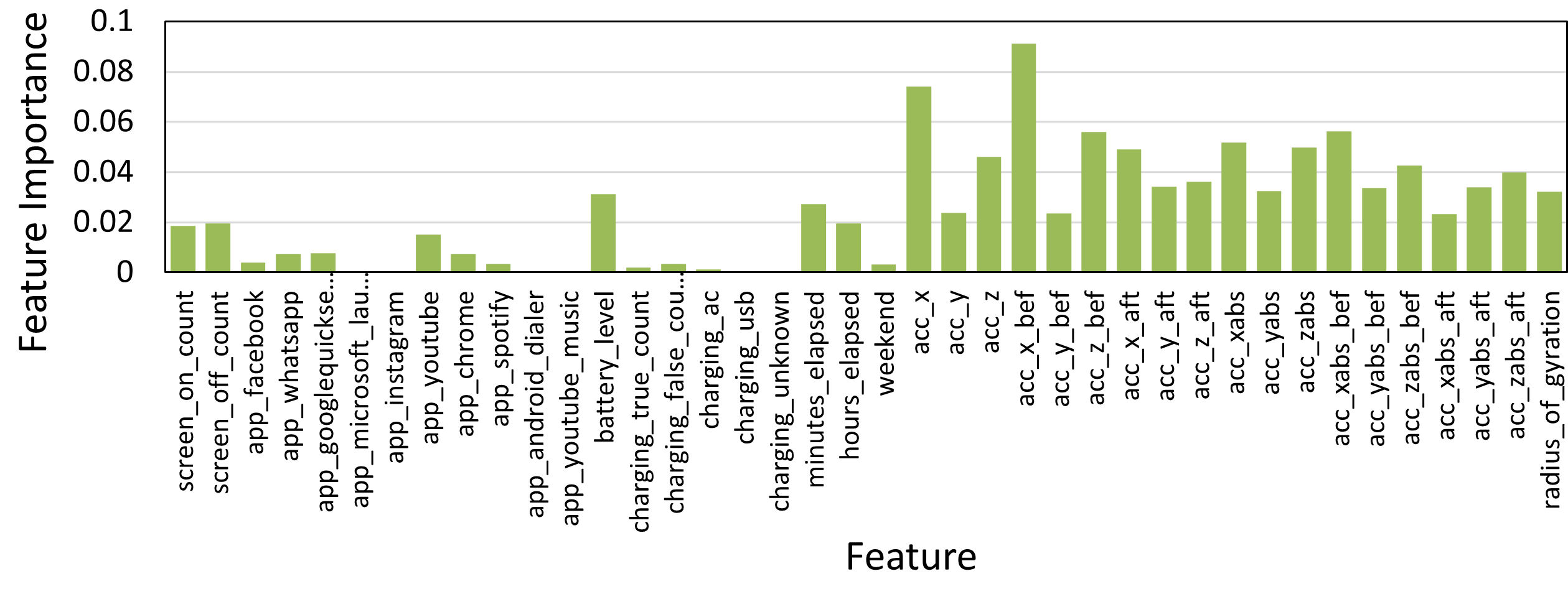}
        \caption{User C}
        \label{fig:fi_user3}
    \end{subfigure}
    
    \caption{Feature importance values from RFs for (a) BASE and (b)--(d) PERS2 models from three randomly selected users.}
    \label{fig:feature_importance_pers2}
\end{center}
\vspace{-0.2 in}
\end{figure*}

In order to study individual differences further, Figure~\ref{fig:feature_importance_pers2} gives BASE feature importance values (Figure~\ref{fig:fi_base}) and PERS2 feature importance values for three randomly selected users (Figure~\ref{fig:fi_user1}, Figure~\ref{fig:fi_user2}, and Figure~\ref{fig:fi_user3}) based on Gini importance of RF classifiers. These figures illustrate how datasets from different users have varying features that could discern between eating and non-eating. For example, BASE (0.081, 0.055) and \textit{user A} (0.076, 0.048) had high values for time of the day (minutes\_elapsed, hours\_elapsed). However, for \textit{user C} (0.027, 0.019) the value for those features are low. Moreover, BASE (0.059) and \textit{user B} (0.053) had high values for battery\_level, whereas for \textit{user A} (0.037) and \textit{user C} (0.031), had lower values compared to other features. Another feature that had a clear difference was radius\_of\_gyration. It was high for BASE (0.058), \textit{user A} (0.062), \textit{user B} (0.049) and low for \textit{user C} (0.031). On a feature group level, the ACC feature group had very similar values across all features [0.035,0.040] in BASE, whereas \textit{user A} and \textit{user C} had highly varied feature values, where some ACC feature values are comparatively higher than other features values. These findings point toward why PERS2 models provided significantly higher accuracy than the BASE. This is because PERS2 capture user-specific behaviors regarding eating and non-eating events.

\begin{table}[t]
        % \small
        % \centering
        
        \caption{Personalized eating event detection accuracy breakdown for random five users in PERS2 with ALL-FS. F1-score (F1) and AUROC are shown. Top performing feature group, and modalities included in the feature group are shown with \checkmark. F1-score Bump indicates the F1-score of BASE \textcolor{cobalt}{(Blue)}, increase in F1-score from BASE to PERS1 \textcolor{green(munsell)}{(Green)}, from PERS1 to PERS2 \textcolor{amber(sae/ece)}{(Orange)} for each user.}
        % \resizebox{0.8\textwidth}{!}{%
        \resizebox{\columnwidth}{!}{%
        \begin{tabular}{l l l l l l l l l l}
        
        %&
        %&
        %& 
        %\multicolumn{1}{r}{\ChartBarBlue{1}}&
        %BASE &
        %\multicolumn{1}{r}{\ChartBarGreen{1}}& 
        %PERS1 &
        %\multicolumn{1}{r}{\ChartBarOrange{1}}&
        %PERS2 &
        %\\
        
        % &
        % &
        % & 
        % &
        % &
        % & 
        % &
        % &
        % &
        % \\

        \rowcolor{gray!15}
        \textbf{User}& 
        \textbf{F1, AUROC} &
        % \textbf{Feature} &
        \textbf{LOC} &
        \textbf{ACC} &
        \textbf{APP} &
        \textbf{BAT} &
        \textbf{SCR} &
        \textbf{TIME} &
        \textbf{F1-score Bump} 
        \\

        \hline 
        
        1 & % c656d
        0.92, 0.82 & % acc, f1, AUROC-roc, kappa
        % U1-F5 & % feature group ['screen_off_count', 'app_facebook', 'app_whatsapp', 'app_googlequicksearchbox', 'app_microsoft_launcher', 'app_instagram', 'app_chrome', 'app_spotify', 'charging_true_count', 'charging_ac', 'charging_usb', 'charging_unknown', 'weekend', 'acc_x_bef', 'acc_y_bef'] 
         & % LOC
        \checkmark & % ACC
        \checkmark & % APP
        \checkmark & % BAT
        \checkmark & % SCR
        \checkmark & % TIME
        \Chart{0.66}{0.08}{0.013} % BASE10 = 89.8  , PERS1 = 92.1
        \\
        
        \rowcolor{gray!5}
        2 & % b8cd9
        0.84, 0.59 & % acc, f1, AUROC-roc, kappa
        % U3-F4 & % feature group ['screen_on_count', 'screen_off_count', 'app_facebook', 'app_whatsapp', 'app_googlequicksearchbox', 'app_microsoft_launcher', 'app_instagram', 'app_youtube', 'app_chrome', 'app_spotify', 'app_android_dialer', 'app_youtube_music', 'battery_level', 'charging_true_count', 'charging_false_count', 'charging_ac', 'charging_usb', 'charging_unknown', 'weekend', 'acc_yabs_bef']
         & % LOC
        \checkmark  & % ACC
        \checkmark  & % APP
        \checkmark  & % BAT
        \checkmark  & % SCR
        \checkmark  & % TIME
        \Chart{0.40}{0.03}{0.05} % BASE10 = 82 , PERS1 = 83
        \\

        3 & % fcf32
        0.88, 0.91 & % acc, f1, AUROC-roc, kappa
        % U21-F2 & % feature group ['screen_on_count', 'app_facebook', 'app_whatsapp', 'app_googlequicksearchbox', 'app_microsoft_launcher', 'app_instagram', 'app_youtube', 'app_chrome', 'app_spotify', 'app_android_dialer', 'app_youtube_music', 'battery_level', 'charging_true_count', 'charging_false_count', 'charging_ac', 'charging_usb', 'charging_unknown', 'weekend', 'acc_x', 'acc_y', 'acc_x_bef', 'acc_z_bef', 'acc_x_aft', 'acc_y_aft', 'acc_z_aft', 'acc_xabs', 'acc_yabs_bef', 'acc_zabs_bef', 'acc_xabs_aft', 'radius_of_gyration']
        \checkmark & % LOC
        \checkmark & % ACC
        \checkmark & % APP
        \checkmark & % BAT
        \checkmark & % SCR
        \checkmark & % TIME
        \Chart{0.03}{0.47}{0.14} % BASE10 = 70 , PERS1 = 84.2
        \\
        
        % 7 & % f2fd5
        % 88.2, 85.5, 0.74, 0.35 & % acc, f1, AUROC-roc, kappa
        % % U26-F5 & % feature group ['screen_on_count', 'app_facebook', 'app_whatsapp', 'app_googlequicksearchbox', 'app_microsoft_launcher', 'app_instagram', 'app_youtube', 'app_chrome', 'app_spotify', 'app_android_dialer', 'app_youtube_music', 'charging_true_count', 'charging_false_count', 'charging_ac', 'weekend']
        %  & % LOC
        %  & % ACC
        % \checkmark & % APP
        % \checkmark & % BAT
        % \checkmark & % SCR
        % \checkmark & % TIME
        % \Chart{0.0}{0.0}{0.0} % BASE10 = 83.0 , PERS1 = 86.4
        % \\
        
        \rowcolor{gray!5}
        4 & % d35e9
        0.81, 0.48 & % acc, f1, AUROC-roc, kappa
        % U34-F6 & % feature group ['screen_off_count', 'app_facebook', 'app_googlequicksearchbox', 'app_microsoft_launcher', 'app_instagram', 'app_youtube', 'app_chrome', 'app_spotify', 'app_android_dialer', 'app_youtube_music']
         & % LOC
         & % ACC
        \checkmark & % APP
         & % BAT
        \checkmark & % SCR
         & % TIME
        \Chart{0.08}{0.19}{0.23} % BASE10 = 72.5 , PERS1 = 78.2
        \\

        % 9 & % e496d
        % 89.4, 89.1, 0.91, 0.72 & % acc, f1, AUROC-roc, kappa
        % % U37-F6 & % feature group ['screen_off_count', 'app_facebook', 'app_whatsapp', 'app_googlequicksearchbox', 'app_microsoft_launcher', 'app_youtube', 'app_chrome', 'acc_x', 'acc_z_bef', 'acc_zabs'] 
        %  & % LOC
        % \checkmark & % ACC
        % \checkmark & % APP
        %  & % BAT
        % \checkmark & % SCR
        %  & % TIME
        % \Chart{0.0}{0.0}{0.0} % BASE10 =  78.7, PERS1 = 88.7
        % \\
        
        % \rowcolor{gray!5}
        5 & % a77d3
        0.98, 0.99 & % acc, f1, AUROC-roc, kappa
        % U39-F3 & % feature group ['screen_on_count', 'screen_off_count', 'app_facebook', 'app_whatsapp', 'app_googlequicksearchbox', 'app_microsoft_launcher', 'app_instagram', 'app_youtube', 'app_chrome', 'app_spotify', 'app_android_dialer', 'app_youtube_music', 'charging_true_count', 'charging_false_count', 'charging_ac', 'charging_usb', 'charging_unknown', 'hours_elapsed', 'acc_x', 'acc_y', 'acc_z', 'acc_z_bef', 'acc_x_aft', 'acc_y_aft', 'acc_xabs_aft']
         & % LOC
        \checkmark & % ACC
        \checkmark & % APP
        \checkmark & % BAT
        \checkmark & % SCR
        \checkmark & % TIME
        \Chart{0.16}{0.57}{0.2333} % BASE10 =  74.7, PERS1 = 91.7
        \\
        
         &
         & % avg (sd) f1
        %  & % feature group
         & % LOC
         & % ACC
         & % APP
         & % BAT
         & % SCR
         & % TIME
        \tiny 70\% \hspace{1.1cm} 100\%
        \\

        \hline 
        
        \end{tabular}}
        \label{tab:inference_pers2_all_users}

\end{table}

\subsection{Effect of Personalization on Individuals}\label{subsec:user_based_results}

Table~\ref{tab:inference_pers2_all_users} shows PERS2 results breakdown for five random users. The maximum F1 score attained by a user here with PERS2 was 0.98. In addition, for five users here and, in fact, for all 58 users for whom PERS2 models were trained, after feature selection, no two users shared the same feature group that resulted in the highest accuracy. For example, \textit{user 3} used features from all feature groups in the model to attain an F1 score of 0.88, whereas \textit{user 4} used a selected set of features from only APP and SCR to attain an F1 score of 0.81. In addition, F1-score bump representation shows that the PERS2 score of different users increased by different amounts compared to BASE and PERS1 results for individual users. For example, for \textit{user 5}, BASE to PERS2 increase was 24\% whereas for \textit{user 1}, the increase was 2.6\%. This shows how the effect of personalization could vary from user to user depending on the chosen features. It is also worth noting that for no user, both F1-score and AUROC decreased when going from BASE to PERS1 to PERS2.

\section{Discussion and Limitations}\label{sec:discussion}

We now discuss implications as well as limitations of our work.

\paragraph{Time Window for Eating Event Detection}
It is worth noting how a chosen time window affects the inference task. Similar to other inferences of human activities in ubicomp research \cite{Biel2018, Meegahapola2021OneMoreBite, Rodriguez2017, meegahapola2021examining, Bae2017}, eating event inference is carried out with a time-window based approach, with fixed or variable frequencies, depending on the application. For example, typical activity recognition algorithms that use time windows of 3-10 seconds could run once every second with overlapping sensor data segments, or run every thirty seconds to generate sparser inferences. In ubicomp work, Bae et al. \cite{Bae2017} used thirty-minute, one-hour, and two-hour time windows for drinking event detection. 
Meegahapola et al. \cite{Meegahapola2021OneMoreBite} used a one-hour time window to detect subjective food consumption level inferences. This research is in line with Bisogni's contextual framework \cite{bisogni2007dimensions}, which not only considers eating/drinking events but the whole context around them. Similarly, the models discussed in this paper could also be run with different time intervals, depending on the use case. For example, with a one-hour time window, if the inference task was performed for \textbf{T$_{anc} = $} 2.15pm at 2.45pm, and run again for \textbf{T$_{anc} = $} 2.17pm at 2.47pm, the granularity would be higher, compared to running the second inference after another hour for \textbf{T$_{anc} = $} 3.15pm at 3.35pm. In addition, as shown in the previous section, this inference could be run for shorter time windows, obtaining reasonable inference accuracies. Even though the one-hour time window performed better for this dataset, shorter time windows might perform better for other datasets. Coming back to the previous example, with a shorter twenty-minute time window, for \textbf{T$_{anc} = $} 2.15 pm, the inference could be done at 2.25 pm, hence reducing the time between \textbf{T$_{anc}$} and inference time. Hence, the time window and inference frequency should be chosen depending on the use case and the available data. Future work could explore how these aspects affect inference performance in more depth. Furthermore, we produced results for different time windows ranging from ten to ninety minutes (not included here due to space limitations). The best average results were obtained for the sixty minute case, closely followed by fifty minutes and seventy minutes. Hence, only the results for the sixty minute case were included throughout the paper. However, as mentioned earlier, different time windows could be used in future work depending on the dataset and features.

\paragraph{Effect of Automatic Inferences on Battery Life}\label{subsec:batterylife} Many prior automatic dietary monitoring inference systems have used wearable devices, which
required the device to run inference models continuously  at fixed or variable intervals \cite{bell2020automatic}. This could have a significant impact on battery life of wearables. In addition, wearable devices rarely work alone as standalone devices. Typical wearables connect to smartphones, and smartphone apps are used to provide feedback and interventions to people. In the case of reminding users to fill in self-reports, users typically need to fill in food diaries on the phone as it is challenging to use a food diary on a wearable device. Hence, maintaining this connection with the smartphone could also affect the battery life of both the phone and the wearable. Eating event detection on the phone would reduce most of these issues. In the results section, we showed that models performed well for ALL, interaction sensing modalities (INTSEN - SCR, APP), and continuous sensing modalities (CONSEN - LOC, BAT, ACC) for both PERS1 and PERS2. While CONSEN would consume far more battery life as it uses accelerometer and location-related features, INTSEN only uses app usage and screen usage, which are just phone usage logs. This makes INTSEN inferences computationally cheaper than those using CONSEN or ALL. However, the set of features available in the dataset for APP and SCR are not rich, and future data collection efforts could look into collecting richer datasets that could further increase performance. Overall, INTSEN could be a low-cost alternative to phone-based eating event tracking, compared to CONSEN and wearable-based tracking. In this sense, INTSEN could be used as a low-power sensing modality to trigger more accurate, high-power CONSEN or ALL-FS models, as discussed in \cite{bell2020automatic}.

\paragraph{Interpretation of Eating Event Detection}\label{subsec:interpretation}
There are many ways in which an eating event can be interpreted. In this paper, an hour period that contains an eating episode is considered as an eating event. This was done with the assumption that there is a time period in which participants prepare for eating (going to the place of eating, preparing food, etc.) and move on to other activities %return to their routine behavior
after the eating event. Hence, the objective was to capture all such behaviors using sensing modalities. This is in line with the idea of capturing holistic food consumption events similar to prior work in smartphone sensing \cite{Biel2018, bisogni2007dimensions, Bae2017, meegahapola2021examining, Meegahapola2021OneMoreBite, meegahapola2021examining}. However, some previous studies only investigated detecting eating gestures from hands, neck, chews, or mastication, and thus in many such cases, eating episodes are detected by sensing different phenomena (e.g., hand motion in wrist-based sensing, chewing in earable-based sensing, or jaw bone movement in necklace-based sensing). In addition, some studies have modeled eating event detection as a time-series data analysis problem. In contrast, the approach discussed in our paper did not consider the time-series nature of the data, and we extracted eating and non-eating events using short-term retrospective self-reports to model inferences using a tabular dataset (Figure~\ref{fig:objective}). Due to these differences in sensing and modeling techniques, 
%there is no inherent definition of an eating event or an eating episode, and 
there is no direct comparison possible with previous work. It is worth noting that there exist subtle differences in how eating events/episodes are defined in different studies, and results should be interpreted with caution by understanding the essence of each individual work.

\paragraph{About to Eat, Eating Now, or Just Ate?}\label{subsec:time_of_eating} Even though previous research has used terms such as eating event detection, eating moment recognition, and eating detection, there is a fundamental difference between {\it what} is being sensed and {\it when} it is being sensed. In addition, depending on when sensing occurs, there are differences w.r.t. how such sensing techniques can be used to benefit users. For example, Rahman et al. \cite{Rahman2016} explicitly mentioned that they are predicting about-to-eat moments to predict eating episodes before they occur. Such inferences are useful for interventions because they can motivate a user not to eat, or used to ask users to control their eating amount before an episode occurs. However, this method might not be used for automated food tracking because predicted eating episodes might or might not happen due to interventions. Furthermore, other studies attempt to detect the episode during the actual eating action \cite{thomaz2015practical, bedri2017earbit, chun2018detecting}. In such cases, it is challenging to ask users not to eat because they are already doing it. However, these approaches could be used for certain interventions and automated food tracking. Additionally, assume that there is a need for users to complete mobile food diaries. In this case, it might be less desirable to trigger reminders at the moment itself, as prior work has shown that people do not appreciate it when they are disturbed during eating moments \cite{Meegahapola2021Survey, Biel2018}. On the other hand, in our approach, an eating event is detected retrospectively, and approximately less than thirty minutes after the end of eating. The proposed technique is less useful for in-the-moment interventions. However, it could be used for interventions regarding future eating events. The proposed technique would also work for automated food tracking, as it would allow reminding users to fill in food diary reports within a short time after eating, hence not disturbing them during actual eating times. This could also reduce recall bias because it would not be too far from the actual eating episode. Hence, as explained, depending on the time events are sensed, the most appropriate use-cases would be different. These aspects should be considered when building future sensing techniques for mobile food diaries.

\paragraph{Other Informative Features}\label{subsec:time_of_eating}
It is important to acknowledge that it was challenging to generate interpretable and meaningful features from the dataset. First, it was clear from the results that ACC features were informative of eating and non-eating events. This means that activity levels are helpful in discerning between the two classes. However, the only available features were statistical features generated from the accelerometer axes, which are hard to interpret. More easily interpretable features such as activity level (i.e., step count) and activity type (i.e., walking, sitting, driving, running, cycling, etc.), which can be captured using activity recognition engines in modern smartphones, were not available in the dataset. Future work should look into capturing such interpretable features, which can be generated with low-power consumption on a smartphone. It was also impossible to determine app usage times for features in the APP group because such features were not available in the dataset. This is another aspect that could be improved when creating mobile sensing applications for future studies. Another challenge in the data filtering and processing phase was missing data, especially from the location sensor. As the location sensor consumes a high amount of power, there is a tendency for participants to turn off this sensor. Finally, researchers could examine other modalities such as touch and typing events, notification clicking behaviors, and continuous sensing modalities such as ambient light sensors, which are typically available in modern smartphones and have shown promise in other smartphone sensing based behavioral modeling tasks \cite{Meegahapola2021Survey}.

\paragraph{Importance of Diversity-Awareness}\label{subsec:time_of_eating}
Depending on demographic attributes, lifestyle, and culture, eating behaviors can significantly vary \cite{rolls1991gender, ma2015food}. Early work documented differences between men and women related to eating behavior \cite{rolls1991gender}.
%found that men tend to ingest more calories while women tend to be in more social settings when eating 
Other statistics show country-based differences. For instance, people in European countries like France, Italy, and Spain spend more time a day on average eating and drinking than people in the United States \cite{McCarthy2020Where}.
%two hours a day eating and drinking, while in the United States, people spend only one-hour \cite{McCarthy2020Where}. 
Our work has studied the eating behavior of a group of college students in Mexico. The results can not be assumed to represent the eating behavior of other age groups from the same country or people from other countries. Prior work has highlighted the importance of considering diversity-awareness in social platforms that use mobile sensing and machine learning \cite{Khwaja2019, schelenz2021theory}. Hence, future work needs to be carried out for different age groups and countries.

\paragraph{Additional Assumptions and Implications}\label{subsec:limitations} 

In the first place, our work assumes that human behavior does not change significantly over time with respect to app usage, screen usage, and activities. However, this is not always the case as the lifestyle and behavior of people could indeed change over a period of time. Future work could look into using time windows (e.g. one month, two months, etc.) when selecting data for training models. Unfortunately, the dataset used in our work did not allow to capture such behavioral changes because the data collection involved only a few weeks. Examining such temporal behavioral change was not a goal of this study and is out of the scope of our paper. 

In the second place, we assumed that there is no relation between eating and non-eating events of the same person, even within the same day. However, this might not be the case because eating and not eating are temporally linked, e.g., long periods of not eating could increase the possibility of eating. On the flip side, eating a meal right now would increase the possibility of non-eating in the next few hours. With the studied dataset, it was not possible to test such phenomena because only three self-reports were collected per day. Therefore, all eating and non-eating events might not be present. This topic is open for future investigation. 

In the third place, another limitation is that eating events could be under-reported for convenience or other reasons (i.e., a person reporting non-eating in a case when an eating event indeed occurred). Even though this might add some noise to the dataset, prior work also suggests that it is difficult to capture and verify all self-reports during real-life experiments \cite{Rodriguez2017, Biel2018, Santani2018, LiKamWa2013}. %Further, a one-hour time window to aggregate sensor data around self-reports is a choice for this particular dataset. However, exploring other time windows is a possibility for future studies. 

As a fourth issue, general machine learning models with feature level fusion were used in our work, similar to many prior smartphone sensing in-the-wild deployments \cite{Biel2018, Santani2018, Rodriguez2017, Bae2017}. Whether other decision-level fusion techniques could be used for this task is open for future investigation. 

Finally, regarding statistical analyses, even though we used a reasonably larger sample size compared to other ubicomp studies, future work in this domain could examine larger sample sizes to examine the issues of generality and personalization in more depth.
%Finally, it is worth noting that the technique proposed in this study might not be suitable for clinical populations for whom more fine-grained techniques would be needed. However, it was not the focus of this study. Hence, future work could look into fine-grained eating episode detection using smartphones. Finally, 
Note also that the Bonferroni correction \cite{Vickerstaff2019} was not used when calculating p-values even though there were multiple comparisons in the statistical analysis. So, the results regarding p-values need to be interpreted with caution.

\section{Conclusion}\label{sec:conclusion}

In this paper, we examined the eating behavior of 58 college students in Mexico using self-reports and passive smartphone sensing data. First, it was shown that time of the day, and features from modalities such as screen usage, accelerometer, app usage, and location are indicative of eating and non-eating events. Then, it was shown that eating and non-eating events can be inferred with an AUROC of 0.65 (F1-score of 0.75) using a subject-independent model, which can be further improved up to AUROC of 0.81 (F1-score of 0.85 for subject-dependent and 0.81 for hybrid models) using personalization techniques. Using feature importance values from classification models and sequential forward feature selection techniques, our work showed that best-performing, subject-dependent models for different users rely on different feature groups. These findings are encouraging towards future mobile food diary apps that are context-aware for both user- and population-level use cases.

\section{Acknowledgement}

We thank William Droz (Idiap Research Institute) for support with data analysis and discussions. We also thank Viridiana del Carmen Robledo-Valero, Emilio Ernesto Hernandez-Huerfano, and Leonardo Alvarez-Rivera (IPICYT) for their previous work with data collection.

\bibliography{12-citations.bib}{}

\begin{thebibliography}{10}
\bibitem{alexandercan}
J.~Alexander, ``Can journaling help you heal from your eating disorder?''
\end{thebibliography}
\bibliographystyle{IEEEtran}

\newpage 

\begin{IEEEbiography}[{\includegraphics[width=1in,height=1.25in,clip,keepaspectratio]{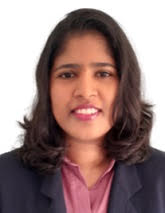}}]{Wageesha Bangamuarachchi} was born in Colombo, Sri Lanka in 1997. She is a final year undergraduate student at University of Moratuwa where she is pursuing a B. Sc. Engineering degree in Computer Science and Engineering. She has worked as a software engineering intern at the Sysco Labs Technologies, Sri Lanka. Her current research interests include ubiquitous computing, mobile sensing, and machine learning.
\end{IEEEbiography}

\begin{IEEEbiography}[{\includegraphics[width=1in,height=1.25in,clip,keepaspectratio]{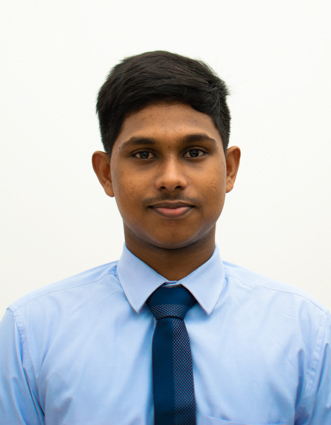}}]{Anju Chamantha} is currently a final year undergraduate studying for his B. Sc. Engineering degree at the Department of Computer Science \& Engineering, University of Moratuwa, Sri Lanka. He has completed his internship at WSO2 Lanka Ltd, a reputed open source software development company. His research interests are in the areas of machine learning, mobile sensing, and human-computer interaction.
\end{IEEEbiography}

\begin{IEEEbiography}[{\includegraphics[width=1in,height=1.25in,clip,keepaspectratio]{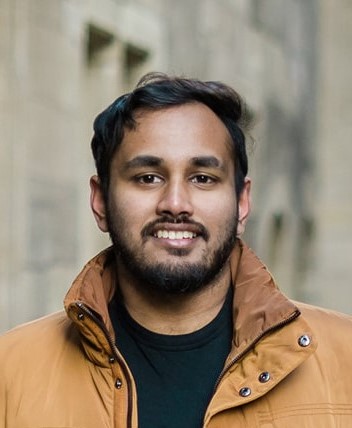}}]{Lakmal Meegahapola} received his B.Sc. in Computer Science and Engineering from the University of Moratuwa, Sri Lanka in 2018. Currently, he is a Doctoral Candidate at the École polytechnique fédérale de Lausanne (EPFL), Switzerland and a Research Assistant in the Social Computing Group at Idiap Research Institute, Switzerland. Lakmal is interested and experienced in using machine learning and data mining techniques for research in the intersection of (a) mobile health sensing, (b) context-awareness, and (c) diversity-aware machine learning.
\end{IEEEbiography}

\begin{IEEEbiography}[{\includegraphics[width=1in,height=1.25in,clip,keepaspectratio]{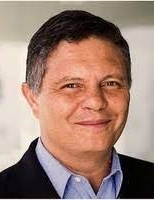}}]{Salvador Ruiz-Correa} received the Ph.D. degree in electrical engineering from the University of Washington. He is currently an Adjunct Professor and a Researcher with the Instituto Potosino de Investigación Científica y Tecnológica, where he leads the Youth Innovation Laboratory (You-i Lab). He co-directs the Center for Mobile Life (Ce Mobili), a research initiative in México. His research interests range from machine learning and computer vision applications to social computing in development contexts, data for social good, and citizen innovation. 
\end{IEEEbiography}

\begin{IEEEbiography}[{\includegraphics[width=1in,height=1.25in,clip,keepaspectratio]{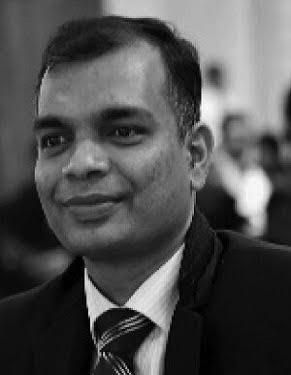}}]{Indika Perera} is a Senior Member of the IEEE and received the BSc Engineering (Hons.) and MSc degrees from the University of Moratuwa, Sri Lanka, the M.B.S. and P.G.D.B.M. degrees from the University of Colombo, Sri Lanka and the PhD degree from the University of St Andrews, U.K. He is currently a Professor and the Head of the Department of Computer Science and Engineering of University of Moratuwa. His research interests include software architecture, software engineering, technology enhanced learning, user experience and application development for bio-health research.
\end{IEEEbiography}

\begin{IEEEbiography}[{\includegraphics[width=1in,height=1.25in,clip,keepaspectratio]{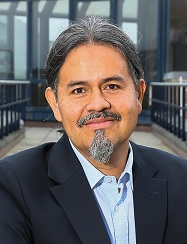}}]{Daniel Gatica-Perez} directs the Social Computing Group at Idiap Research Institute, and is also a Professor at EPFL, Switzerland. His research interests span social computing, ubiquitous computing, and crowdsourcing for social good. He is a member of the IEEE. 
\end{IEEEbiography}

\section{Appendix}
\appendix

\begin{table*}[h]
        \small
        \centering
        \caption{Feature Groups Used in Different Inference Models}
        \resizebox{\textwidth}{!}{%
        \begin{tabular}{l l l}

        \rowcolor{gray!15}
        \textbf{Table}& 
        \textbf{Feature Group} &
        \textbf{Features} 
        \\

        \hline

        Table~\ref{tab:inference_results} &
        F1 & 
        screen\_on\_count, screen\_off\_count, facebook, whatsapp, googlequicksearchbox,
        \\
         &
         & 
        microsoft\_launcher, instagram, youtube, chrome, spotify, android\_dialer, youtube\_music, 
        \\
         &
         & 
        battery\_level, charging\_true\_count, charging\_false\_count, charging\_ac, charging\_usb,
        \\
         &
         & 
         charging\_unknown, minutes\_elapsed, hours\_elapsed, weekend, acc\_x\_bef, acc\_y\_bef, 
        \\
         &
         & 
         acc\_z\_bef, acc\_x\_aft, acc\_y\_aft, acc\_z\_aft, acc\_yabs, acc\_zabs, acc\_xabs\_bef, acc\_yabs\_bef,
         \\
         &
         & 
         acc\_xabs\_aft, acc\_yabs\_aft, acc\_zabs\_aft, radius\_of\_gyration
        \\

        \rowcolor{gray!5}
         &
        F3 &
        screen\_on\_count, screen\_off\_count, facebook, whatsapp, googlequicksearchbox,
        \\
        \rowcolor{gray!5}
         &
         & 
        microsoft\_launcher, instagram, youtube, chrome, spotify, android\_dialer, youtube\_music, 
        \\
        \rowcolor{gray!5}
         &
         & 
        charging\_true\_count, charging\_false\_count, charging\_ac, charging\_usb, charging\_unknown,
        \\
        \rowcolor{gray!5}
         &
         & 
         minutes\_elapsed, hours\_elapsed, weekend, acc\_z\_bef, acc\_x\_aft, acc\_z\_aft, acc\_yabs, 
        \\
        \rowcolor{gray!5}
        &
        &
        radius\_of\_gyration
        \\

         &
        F7 &
        googlequicksearchbox, microsoft\_launcher, instagram, youtube, charging\_false\_count
        \\

        \rowcolor{gray!5}
        Table~\ref{tab:inference_results_pers} &
        F1 & 
        screen\_on\_count, screen\_off\_count, facebook, whatsapp, googlequicksearchbox,
        \\
        \rowcolor{gray!5}
         &
         & 
        microsoft\_launcher, instagram, youtube, chrome, spotify, android\_dialer, youtube\_music, 
        \\
        \rowcolor{gray!5}
         &
         & 
        battery\_level, charging\_true\_count, charging\_false\_count, charging\_ac, charging\_usb,
        \\
        \rowcolor{gray!5}
         &
         & 
         charging\_unknown, minutes\_elapsed, hours\_elapsed, weekend, acc\_x\_bef, acc\_y\_bef, 
        \\
        \rowcolor{gray!5}
         &
         & 
         acc\_z\_bef, acc\_x\_aft, acc\_y\_aft, acc\_z\_aft, acc\_yabs, acc\_zabs, acc\_xabs\_bef, acc\_yabs\_bef,
         \\
         \rowcolor{gray!5}
         &
         & 
         acc\_xabs\_aft, acc\_yabs\_aft, acc\_zabs\_aft, radius\_of\_gyration
        \\

        \hline 
        
        \end{tabular}}
        \label{tab:appendix_feature_groups}

\end{table*}

%[F1] screen\_on\_count, screen\_off\_count, facebook, whatsapp, googlequicksearchbox, microsoft\_launcher, instagram, youtube, chrome, spotify, android\_dialer, youtube\_music, battery\_level, charging\_true\_count, charging\_false\_count, charging\_ac, charging\_usb, charging\_unknown minutes\_elapsed, hours\_elapsed, weekend, acc\_x\_bef, acc\_y\_bef, acc\_z\_bef, acc\_x\_aft, acc\_y\_aft, acc\_z\_aft, acc\_yabs, acc\_zabs, acc\_xabs\_bef, acc\_yabs\_bef, acc\_xabs\_aft, acc\_yabs\_aft, acc\_zabs\_aft, radius\_of\_gyration

% [F3] screen\_on\_count, screen\_off\_count, facebook, whatsapp, googlequicksearchbox, microsoft\_launcher, instagram, youtube, chrome, spotify, android\_dialer, youtube\_music, charging\_true\_count, charging\_false\_count, charging\_ac, charging\_usb, charging\_unknown, minutes\_elapsed, hours\_elapsed, weekend, acc\_z\_bef, acc\_x\_aft, acc\_z\_aft, acc\_yabs, radius\_of\_gyration

% [F7] googlequicksearchbox, microsoft\_launcher, instagram, youtube, charging\_false\_count

\EOD

\end{document}